\documentclass[%
reprint,
superscriptaddress,
 amsmath,amssymb,
 aps,
]{revtex4-2}

\usepackage{graphicx,setspace}% Include figure files
\usepackage{dcolumn}% Align table columns on decimal point
\usepackage{bm}% bold math
\usepackage[mathlines]{lineno}% Enable numbering of text and display math
\usepackage[colorlinks=true,citecolor=teal,linkcolor=cyan,anchorcolor=green,urlcolor=violet]{hyperref}
\usepackage{mathrsfs}
\usepackage{xcolor}
\usepackage{float}
\usepackage[normalem]{ulem}
\usepackage[utf8]{inputenc}
\usepackage{tensor}
\usepackage{physics}
\usepackage{blindtext}
\usepackage{slashed}
\usepackage{ulem}

\newcommand{\be}{\begin{equation}}
\newcommand{\ee}{\end{equation}}
\newcommand{\ba}{\begin{eqnarray}}
\newcommand{\ea}{\end{eqnarray}}

\begin{document}

\title{Conformal scalar NUT-like dyons in conformal electrodynamics}
\author{Ming Zhang}
\email{mingzhang@jxnu.edu.cn}
\affiliation{Department of Physics, Jiangxi Normal University, Nanchang 330022, China}
\author{Jie Jiang}
\email{Corresponding author. jiejiang@mail.bnu.edu.cn}
\affiliation{College of Education for the Future, Beijing Normal University, Zhuhai 519087, China}
\date{\today}

\begin{abstract}
We construct conformally scalar NUTty dyon black holes in non-linear electrodynamics (NLE) which possess conformal and duality-rotation symmetries and are characterized by a free dimensionless parameter. The thermodynamic first law of the black hole is formulated. Then we explore the strong gravitational effects of the black hole, mainly focusing on the innermost stable circular orbits (ISCOs) of massive particles and shadows formed by the photons. One of our interesting results is that if the black hole is endowed with a positive real conformal scalar charge rendering it Reissner-Nordström-like, the radii of the ISCOs and the shadow both increase with the increasing NLE parameter signifying the increasing nonlinearity of the electromagnetic field.
\end{abstract}
\maketitle

\section{Introduction}
In the light scattering process, nonlinearities of the electromagnetic interaction are discovered and experimentally proved \cite{ATLAS:2017fur}. The exploration of the nonlinear electrodynamics (NLE) theories in general relativity (GR) theory was inspired by solving the singularity problem of a black hole. Usually the NLE theories are derived from the Lagrangians constructed with quadratic electromagnetic invariants $F_{\mu\nu} F^{\mu\nu}$ and $F_{\mu\nu} { }^{*}F^{\mu\nu}$. There is NLE depending only on the invariant $F_{\mu\nu} F^{\mu\nu}$ proposed by Max Born in \cite{Born:1934ji}, albeit the one depending on both invariants proposed in \cite{Born:1934gh} known as the Born--Infeld theory, which preserves Maxwell equations' electromagnetic duality invariance, seems to be more prevalent. Moreover, a one-loop QED correction to Maxwell’s theory incorporating vacuum polarization effects was raised by Heisenberg and Euler \cite{Heisenberg:1936nmg}. However, none of them keep the conformal invariance characteristics which is endowed to Maxwell theory.  Recently, a nonlinear extension of the source-free Maxwell electrodynamics which preserves both conformal invariance and $SO(2)$ electromagnetic duality was proposed in \cite{Bandos:2020jsw,Kosyakov:2020wxv}. Following the convention used in \cite{BallonBordo:2020jtw}, we in this paper call it the {\it{conformal electrodynamics}}, though in \cite{Bandos:2020jsw,Flores-Alfonso:2020euz,Bokulic:2021dtz} it is also named as {\it{ModMax theory}}. The conformal electrodynamics is characterized by a dimensionless parameter $\gamma$, with the Maxwell theory being recovered in the $\gamma\to 0$ limit. For $\gamma>0$ the polarization mode is subluminal and for $\gamma<0$ the polarization mode is superluminal, so in this paper we consider the former case.

The no-hair theorem states that a black hole holds no more hairs other than mass, electric charge, and angular momentum \cite{Ruffini:1971bza}. A minimally coupled scalar cannot be held in the Einstein-scalar theory \cite{Herdeiro:2015waa}, however, being as a counterexample, the Bocharova-Bronnikov-Melnikov-Bekenstein (BBMB) black hole in the Einstein-conformal scalar vacuum leads to a scalar hair \cite{bocharova1970exact} that being unbounded at the horizon but not physically troublesome \cite{bekenstein1975black}. The Einstein-Maxwell-conformally coupled scalar (EMCS) theory triggered much attention in community \cite{Anabalon:2012tu,Simovic:2020dke,Zou:2020zxq}, which, besides the BBMB black hole, also gives a three-dimensional black hole with the scalar field being regular everywhere \cite{Martinez:1996gn} and a scalar hairy black hole with a constant scalar field \cite{Astorino:2013sfa}. For the latter one, emission rate of charged particles was investigated in \cite{Chowdhury:2019uwi}, the weak cosmic censorship conjecture was tested in \cite{Jiang:2020btc}, and it was turned out to be stable against full perturbations \cite{Chowdhury:2018pre,Zou:2019ays}.

Recently, soon after the proposition of the conformal electrodynamics, its applications in the Einstein gravity were put forward \cite{BallonBordo:2020jtw,Flores-Alfonso:2020euz}. It was shown that there is a screening factor that shields the actual charges of the black hole, and the electric and magnetic fields change qualitatively comparing to the Einstein-Maxwell counterpart. In this paper, we will seek Taub-NUT-like black hole solution in the Einstein-conformal electromagnetic system with a conformally coupled scalar field, which substitutes the Maxwell field in the EMCS with the conformal electrodynamic field. The reason we choose the Taub-NUT-like spacetime is that it provides a representative candidate for which the magnetic fields emergent and the conformal electrodynamics becomes non-trivial. We want to investigate how the conformal electrodynamics and the conformally coupled scalar field interact to modify the Einstein gravity. One may also wonder how the conformal electrodynamics deviates from linear electrodynamics \cite{BallonBordo:2020jtw}, in the situation where the conformally coupled scalar field is added. Thus we will show the strong gravitational effect of the solution, by investigating the innermost stable circular orbits (ISCOs) of charged massive particles around the black hole as well as studying the shadow of the object. To these ends, the NUTty dyons solution together with its thermodynamics will be given in Sec. \ref{sec1}. The gravitational effects of the black hole will be elaborated in Sec. \ref{sec2} where the ISCOs of charged massive particles will be studied and in Sec. \ref{sec3} where the shadow of the black hole will be explored. Throughout this paper the units are chosen to be ${\rm{c}}={\rm{\hbar}}={\rm{G}}=1$. Notice that in this paper ${\rm{e}}$ denotes the natural constant and $e$ is the electric charge parameter.

\section{Conformally scalar NUTty dyons solution in conformal electrodynamics}\label{sec1}

\subsection{Solution}
The theory we consider consists of the Einstein gravity, conformally coupled scalar field, and conformal electrodynamics, whose bulk action reads
\begin{equation}\label{act}
I=\frac{}{} \int {\rm{d}}^{4} x \mathcal{L}=I_{{\rm{G}}}+I_{{\rm{CS}}}+I_{{\rm{CE}}},
\end{equation}
where
\begin{equation}
I_{{\rm{G}}}=\frac{1}{2 \kappa} \int {\rm{d}}^{4} x \sqrt{-g}R,
\end{equation}
\begin{equation}
I_{{\rm{CS}}}=-\frac{1}{2} \int {\rm{d}}^{4} x \sqrt{-g}\left[g^{\mu \nu} \nabla_{\mu} \Psi \nabla_{\nu} \Psi+\xi_{D} R\Psi^{2}\right],
\end{equation}
\begin{equation}
I_{{\rm{CE}}}=-\frac{1}{4\pi}\int {\rm{d}}^4 x\sqrt{-g}\mathcal{L}_{CE},
\end{equation}
with
\begin{equation}
\mathcal{L}_{{\rm{CE}}}=-\frac{1}{2}\left(\mathcal{S} \cosh \gamma-\sqrt{\mathcal{S}^{2}+\mathcal{P}^{2}} \sinh \gamma\right),
\end{equation}
\begin{equation}
\mathcal{S}\equiv\frac{1}{2} F_{\mu \nu} F^{\mu \nu}, \quad \mathcal{P}\equiv\frac{1}{2} F_{\mu \nu}({ }^{*}F)^{\mu \nu},
\end{equation}
$\kappa=8\pi$, $R$ is the Ricci scalar, $\Psi$ is the conformally coupled scalar field, the electromagnetic field strength $F_{\mu \nu}$ is given by $F_{\mu \nu}=\nabla_{\mu} A_{v}-\nabla_{\nu} A_{\mu}$, with $A_\mu$ the vector potential. $\mathcal{S}$ and $\mathcal{P}$ are gauge-invariant Lorentz electromagnetic field invariants, which in the Minkowski spacetime are both zero. $\gamma$ is a dimensionless parameter characterizing the NLE. When $\gamma=0$, $L_{{\rm{CE}}}$ reduces to the Maxwell theory, when $\gamma$ increases, we can deem that the extent of the NLE's deviation from the Maxwell theory also increases. The value of $\xi_D$ is chosen to be $\xi_{D}=(D-2) /(4D-4)$ with $D=4$ the spacetime dimensions, such that $I_{{\rm{CS}}}$ together with the equation of motion for the scalar field is invariant under the conformal transformations
\begin{equation}\label{ctmet}
g_{\mu \nu} \rightarrow \Omega^{2} g_{\mu \nu}, \Psi \rightarrow \Omega^{1-D / 2} \Psi,
\end{equation}
with $\Omega$ a transformation function, and this is the reason of the scalar being conformally coupled, though the full action is not necessarily conformal invariant \cite{Martinez:1996gn,Cunha:2016bpi}.

$\mathcal{L}_{{\rm{CE}}}$ possesses both $SO(2)$ duality-rotation (or electromagnetic duality) invariance and conformal invariance and it is a generalization of the Maxwell theory. To see this, we can see the Euler–Lagrange equation and the Bianchi identity
\begin{equation}\label{feq1}
\nabla_{\mu} E^{\mu \nu}=0,
\end{equation}
\begin{equation}\label{feq2}
\nabla_{\mu}{ }^{*} F^{\mu \nu}=0,
\end{equation}
where the strength tensor is defined by
\begin{equation}
E_{\mu v}=\frac{\partial \mathcal{L}}{\partial F^{\mu \nu}}=2\left(\mathcal{L}_{\mathcal{S}} F_{\mu \nu}+\mathcal{L}_{\mathcal{P}}^{*} F_{\mu \nu}\right),
\end{equation}
with 
\begin{equation}
{\cal L_S}= \pdv{\mathcal{L}}{\mathcal{S}}
=\frac{1}{2}\left(\frac{\cal S}{\sqrt{\mathcal{S}^2 + \mathcal{P}^2}}\sinh \gamma-\cosh\gamma\right),
\end{equation}
\begin{equation}
{\cal L_P}=\pdv{\mathcal{L}}{\mathcal{P}}
=\frac{1}{2}\frac{\cal P}{\sqrt{\mathcal{S}^2 + \mathcal{P}^2}}\sinh \gamma\,.
\end{equation}

Under the electromagnetic duality rotation, we have 
\begin{equation}
E_{\mu \nu}^{\prime}=E_{\mu \nu} \cos \theta+{ }^{*} F_{\mu \nu} \sin \theta,
\end{equation}
\begin{equation}
{ }^{*} F_{\mu \nu}^{\prime}={ }^{*} F_{\mu \nu} \cos \theta-E_{\mu \nu} \sin \theta,
\end{equation}
which mean that $(E_{\mu \nu}^{\prime},\,{ }^{*} F_{\mu \nu}^{\prime})$ is invariant under $SO(2)$ rotation. On the other hand, under the conformal transformation (\ref{ctmet}), the field equations (\ref{feq1}) and (\ref{feq2}) are also invariant, as ${ }^{*}F\to { }^{*}F$ and $E\to E$.

Varying the action (\ref{act}) individually with respect to the metric $g^{\mu\nu}$, the scalar field $\Psi$, we obtain
\begin{equation}\label{eosmet}
R_{\mu \nu}-\frac{R}{2} g_{\mu \nu}=\kappa\left(T_{\mu \nu}^{(S)}+T_{\mu \nu}^{(E M)}\right),
\end{equation}
\begin{equation}\label{eossc}
\square \Psi-\frac{1}{6} R \Psi=0,
\end{equation}
where we denoted $\square=\nabla_\mu\nabla^\mu$, the energy-momentum tensor of the scalar field is
\begin{equation}
\begin{aligned}
T_{\mu \nu}^{(S)}=&\nabla_{\mu} \Psi \nabla_{\nu} \Psi-\frac{1}{2} g_{\mu \nu} \nabla_{\sigma} \Psi \nabla^{\sigma} \Psi\\&+\frac{1}{6}\left[g_{\mu \nu} \square-\nabla_{\mu} \nabla_{\nu}+G_{\mu \nu}\right] \Psi^{2},
\end{aligned}
\end{equation}
and the traceless stress-energy tensor of the conformal electromagnetic field is 
\begin{equation}
\begin{aligned}
T_{\mu\nu}^{{\rm{(EM)}}}&=-\frac{1}{4\pi}\left(2F_{\mu\sigma}F_{\nu}{}^\sigma {\cal L_S}+{\cal P}{\cal L_P} g_{\mu\nu}-{\cal L}g_{\mu\nu}\right)\\&=\frac{1}{4\pi}\left({\cal S} g_{\mu\nu}-2F_{\mu\sigma}F_{\nu}{}^\sigma \right){\cal L_S},
\end{aligned}
\end{equation}
where the criterion for conformal invariance
\begin{equation}
\mathcal{L}_{\mathcal{S}} \mathcal{S}+\mathcal{L}_{\mathcal{P}} \mathcal{P}=\mathcal{L}
\end{equation}
was used in the second step \cite{kosyakov2007introduction}.

We assume the metric as
\be\label{metric}
  ds^2 ={ -f\left[{\rm{d}}t + 2n \cos\theta {\rm{d}} \phi\right]^2}+\frac{{\rm{d}}r^2}{f} + (r^2+n^2){\rm{d}} \Omega_2^{2}\,,\ 
\ee
with $f=f(r)$ the blackening factor, $n$ the NUT parameter, ${\rm{d}}\Omega_2^2$ the metric on the unit sphere, and the electromagnetic potential
\be\label{epa}
A = a\left({\rm{d}}t + 2n \cos\theta {\rm{d}}\phi\right)\,, 
\ee
where $a=a(r)$. We will seek for solutions of $f(r)$ and $a(r)$ in the theory described by the action (\ref{act}), which gives the equations of motion (\ref{feq1}) and (\ref{feq2}) for the conformal electromagnetic field, as well as the ones for the spacetime and the scalar field in (\ref{eosmet}) and (\ref{eossc}). 

For the electromagnetic field under the spacetime ansatz, we can obtain the following quantities
\begin{equation}
 F = -a' dt\wedge {\rm{d}}r + 2na' \cos\theta {\rm{d}}r\wedge {\rm{d}}\phi 
  - 2n a\sin\theta {\rm{d}}\theta\wedge {\rm{d}}\phi,
\end{equation}
\begin{equation}
\begin{aligned}
{ }^{*}F= &{-}\frac{2n a}{n^2+r^2}{\rm{d}}t\wedge {\rm{d}}r {+}\frac{4n^2a}{n^2 + r^2} \cos\theta {\rm{d}}r\wedge {\rm{d}}\phi\\&+(n^2 + r^2)a' \sin\theta {\rm{d}}\theta \wedge {\rm{d}}\phi,
\end{aligned}
\end{equation}
\begin{equation}
 {\cal S}= -{ a'^2}+\frac{4 n^2 a^2}{(n^2+r^2)^2},
\end{equation}
\begin{equation}
{\cal P} = {-}\frac{4 n a a'}{n^2+r^2},
\end{equation}
\begin{equation}
\begin{aligned}
 E =&{a' {\rm{e}}^{\gamma}}{\rm{d}}t\wedge {\rm{d}}r {-} 2n a'{\rm{e}}^\gamma \cos\theta {\rm{d}}r\wedge d\phi\\\quad&+2n a {\rm{e}}^{-\gamma} \sin\theta {\rm{d}}\theta \wedge {\rm{d}}\phi\,,
 \end{aligned}
\end{equation}
\begin{equation}
\begin{aligned}
 { }^{*}E =& {\frac{2an {\rm{e}}^{-\gamma}}{n^2+r^2} {\rm{d}}t\wedge {\rm{d}}r -\frac{4n^2a {\rm{e}}^{-\gamma}}{n^2+r^2}\cos\theta {\rm{d}}r\wedge d\phi}\\ \quad&{-(n^2+r^2)a'{\rm{e}}^\gamma \sin\theta} {\rm{d}}\theta \wedge {\rm{d}}\phi\,,
 \end{aligned}
\end{equation}
where the ${}^{\prime}$ denotes derivative with respect to $r$. Then the field equation gives
\begin{align}
-{\rm{e}}^{\gamma } \Bigl[\left(n^2+r^2\right) a''+2 r a'\Bigr]-\frac{4 {\rm{e}}^{-\gamma } n^2 a}{n^2+r^2}=0\,,
\end{align}
from which we have the specific expressions of $a$,
\begin{equation}
  a(r) = c_1 \sin\left(2{\rm{e}}^{-\gamma} \arctan\frac{r}{n}\right)+c_2 \cos\left(2{\rm{e}}^{-\gamma} \arctan\frac{r}{n}\right),
\end{equation}
where $c_1$ and $c_2$ are integral constants restricted by the asymptotic conditions
\be\label{asye}
\lim_{r\to\infty} q_e = e= \frac{1}{4\pi} \int_{\infty}{ }^{*}E,
\ee
\be\label{asym}
 \lim_{r\to\infty} q_m = -2n g=\frac{1}{4\pi}\int_{\infty} F,
\ee
meaning that the asymptotic electric charge and magnetic charge are $e$ and $-2n g$, respectively. As a result, we get the values of the constants $c_1$ and $c_2$, dependent on the asymptotic charges, as
\begin{align}
  c_1 &= -g \cos\left(2 {\rm{e}}^{-\gamma}\pi\right)- \frac{ e \cos(2 {\rm{e}}^{-\gamma}\pi)}{2n}\,,\\
  c_2 &= -g \sin\left(2 {\rm{e}}^{-\gamma}\pi\right)+ \frac{ e \sin(2 {\rm{e}}^{-\gamma}\pi)}{2n}\,.
\end{align}
Thus, the electromagnetic gauge potential is
\begin{equation}
\begin{aligned}
 a =& -g\cos\Bigl[ {\rm{e}}^{-\gamma}\left(\pi-2\arctan\frac{r}{n}\right)\Bigr]\\&- \frac{e}{2n} \sin\Bigl[ {\rm{e}}^{-\gamma} \left(\pi - 2\arctan\frac{r}{n}\right)\Bigr]\,.
 \end{aligned}
\end{equation}

From the Eq. (\ref{eossc}), we have
\begin{equation}
\frac{3}{2} \Psi\left[\frac{2 \left(-1+f+2 r f^{\prime}\right)}{n^{2}+r^{2}}+f^{\prime \prime}\right]=0,
\end{equation}
for which either $\Psi=0$ or
\begin{equation}
f=1+\frac{c_{3}}{n^{2}+r^{2}}+\frac{r c_{4}}{n^{2}+r^{2}}
\end{equation}
solves it. As the former solution is trivial, we just consider the latter one. The specific values of $c_3$ and $c_4$ can be restricted by the Eq. (\ref{eosmet}), which just yields
\begin{equation}\label{blf}
f(r)=\frac{r^{2}-n^{2}}{n^{2}+r^{2}}+\frac{{\rm{e}}^{-\gamma}\left(e^{2}+4 g^{2} n^{2}\right)\left(e^{2}+\alpha\right)}{e^{2}\left(n^{2}+r^{2}\right)}-\frac{2 m r}{n^{2}+r^{2}},
\end{equation}
\begin{equation}
\Psi=\sqrt{\frac{\alpha}{12\pi\left(e^{2}+\alpha\right)}},
\end{equation}
where $\alpha$ is the conformal scalar parameter, rendering the scalar field $\Psi$ being constant. This scalar hair does not vanish even when the electric charge is absent. Notice that in this paper we will only consider the real scalar field, so that $\alpha>0$ or $\alpha<-e^2$. It is obvious that in the former parameter range, the black hole tends to be Reissner-Nordström-like, while for the latter one, the black hole tends to be Schwarzschild-like, and in the $\gamma\to 0$ and $n\to 0$ limit, this solution reduces to the one in the Maxwell case obtained in \cite{Astorino:2013sfa}.

\subsection{Cohomogeneity Thermodynamics}
Thermodynamics of the black hole with NUT charge have been studied recently in \cite{Mann:2020wad,Abbasvandi:2021nyv,Hennigar:2019ive,Bordo:2019slw,BallonBordo:2020mcs,BallonBordo:2019vrn,Awad:2020dhy}, especially in \cite{BallonBordo:2020jtw} for the Taub-NUT solution in Einstein case with conformal electrodynamics, which are main references for our study here. The event horizon of the NUTty dyon black hole generated by the Killing vector $\xi=\partial_t$ is
\begin{equation}
r_+ =m+\frac{\sqrt{{\rm{e}}^{\gamma}\left(m^{2}+n^{2}\right)-\alpha-4 g^{2} n^{2}-e^{2}-4 g^{2} n^{2} \alpha/e^{2}}}{{\rm{e}}^{\gamma / 2}}.
\end{equation}
The temperature, entropy, and mass of the black hole can be obtained as
\begin{equation}
\begin{aligned}
T&=\left.\frac{1}{4\pi}\frac{{\rm{d}}f(r)}{{\rm{d}}r}\right|_{r=r_+}\\&=\frac{1}{4 \pi r_{+}}\left[1-\frac{{\rm{e}}^{-\gamma}\left(4 g^{2} n^{2}+\alpha\right)}{n^{2}+r_{+}^{2}}-\frac{{\rm{e}}^{-\gamma}\left(e^{4}+4 g^{2} n^{2} \alpha\right)}{e^{2}\left(n^{2}+r_{+}^{2}\right)}\right],
\end{aligned}
\end{equation}
\begin{equation}
S=-2 \pi \oint {\rm{d}}^{2} x \sqrt{\hat{h}} \frac{\partial \mathcal{L}}{\partial R_{a b c d}} \hat{\epsilon}_{a b} \hat{\epsilon}_{c d}=\frac{\pi e^{2}(r_+^2+n^2)}{e^{2}+\alpha},
\end{equation}
\begin{equation}
M=\frac{e^2 m}{e^2+\alpha},
\end{equation}
where $\hat{\epsilon}_{a b}$ is a normal bivector which satisfies $\epsilon_{a b} \epsilon^{a b}=-2$, $\hat{h}$ is the determinant of the induced line element from $g_{\mu\nu}$ at the hypersurface $t={\rm{const.}}$ and $r=r_{+}$, the mass can be obtained by the Euclidean method \cite{Martinez:1996gn,Ashtekar:2003jh}, as we will show in what follows.

As mentioned above in Eqs. (\ref{asye}) and (\ref{asym}), the asymptotic electric and magnetic charges of the black hole are
\begin{equation}
Q=e, \quad Q_{m}=-2 g n,
\end{equation}
 and they are related by the electromagnetic duality
 \begin{equation}\label{duality}
e \leftrightarrow-2 n g, \quad 2 n g \leftrightarrow e.
\end{equation}
At the event horizon, the charges become
\begin{equation}
Q_e^{+}=q_{e}\left(r_{+}\right)=e^{\gamma}\left(n^{2}+r_{+}^{2}\right) a^{\prime}\left(r_{+}\right),
\end{equation}
\begin{equation}
Q_{m}^{+}=q_{m}\left(r_{+}\right)=2 n a\left(r_{+}\right).
\end{equation}
The gauge electric potential can be calculated by extracting the Killing vector with the vector potential as
\begin{equation}
\begin{aligned}
\varphi&=-\left(\left.\xi_{\mu} A^{\mu}\right|_{r=r_{+}}-\left.\xi_{\mu} A^{\mu}\right|_{r\to\infty}\right)\\&=-a\left(r_{+}\right)-g \\
&=g\left[\cos \left({\rm{e}}^{-\gamma}\left(\pi-2 \arctan \frac{r_{+}}{n}\right)\right)-1\right] \\
&\quad+\frac{e}{2 n} \sin \left[{\rm{e}}^{-\gamma}\left(\pi-2 \arctan \frac{r_{+}}{n}\right)\right] .
\end{aligned}
\end{equation}
The magnetic potential 
\begin{equation}
\begin{aligned}
\varphi_{m}=& \frac{e}{2 n}\left(\cos \left({\rm{e}}^{-\gamma}\left(\pi-2 \arctan \frac{r_{+}}{n}\right)\right)-1\right) \\
&-g \sin \left({\rm{e}}^{-\gamma}\left(\pi-2 \arctan \frac{r_{+}}{n}\right)\right)
\end{aligned}
\end{equation}
can be yielded directly based on the electric potential by using the electromagnetic duality (\ref{duality}).

The Gibbs free energy can be obtained by the Euclidean action \cite{Lee:2018hrd,Bueno:2018xqc,Sebastiani:2017rxr,Monteiro:2009tc,Mann:2020wad}
\begin{equation}
\begin{aligned}
\mathcal{I}=& I+I_{\mathrm{GH}} \\
=&- \frac{1}{16 \pi} \int_{M} \mathrm{~d}^{4} x \sqrt{-g}R \\
&-\frac{1}{4 \pi} \int \mathrm{d}^{4} x \sqrt{-g} \mathcal{L}_{{\rm{CE}}}\\& -\frac{1}{8 \pi} \int_{\partial M} \mathrm{~d}^{3} x \sqrt{-h}(1-8\pi\xi_D\Psi^2)\left(K-K_0\right),
\end{aligned}
\end{equation}
where $K_0=2/r$ is the extrinsic curvature of the background flat spacetime. Notice that the Wick rotations $n\to -i n,\,e\to -i e,\,g\to -i g$ (or $n\to i n,\,e\to i e,\,g\to i g$) should be conducted to calculate the action and finally the reverse procedure should also be done (For the $\mathcal{L}_{{\rm{CE}}}$ term, one can first directly calculate $\mathcal{L}_{{\rm{CE}}}$, then do Wick rotation to conduct the integral, and finally rotate back). Then we have the specific expression for the Gibbs energy,
\begin{equation}
\begin{aligned}
G&=\mathcal{I}/\beta\\&=\frac{e\left(e^{2} g+e m+g \alpha\right)}{2\left(e^{2}+\alpha\right)}\\&\quad-\frac{1}{2} eg \cos \left[2 {\rm{e}}^{-\gamma}\left(\pi-2 \operatorname{arctan}\frac{r_+}{n}\right)\right]\\&\quad+\frac{1}{8} \left(4 g^2 n-\frac{e^2}{n}\right) \sin \left[2 e^{-\gamma } \left(\pi -2 \arctan\frac{r_+}{n}\right)\right],
\end{aligned}
\end{equation}
where $\beta$ is the inverse of the temperature. The Gibbs function satisfies
\begin{equation}
\begin{aligned}
G=M-T S-\varphi Q-\psi N,
\end{aligned}
\end{equation}
where $N$ is the Misner charge conjugated to the Misner potential $\psi$. The conformal scalar, though being a primary hair, here will not enter the first law of the black hole, which reads
\begin{equation}
\delta G=-S \delta T-N \delta \psi-Q \delta \varphi+\varphi_{m} \delta Q_{m}^{+}.
\end{equation}

After taking the Misner potential $\psi$ as 
\begin{equation}
\psi=\frac{\kappa_\pm}{4\pi}=\frac{1}{8\pi n},
\end{equation}
where $\kappa_\pm$ are surface gravity corresponding to the Killing vectors
\begin{equation}
k_\pm=\partial_t\pm\frac{1}{2n}\partial_\phi,
\end{equation}
the integration Smarr relation for the black hole then can be written as
\begin{equation}\label{smax}
M=2 T S+\phi Q+\phi_{m} Q_{m}^{+}+2 \psi N.
\end{equation}

Note that $\psi$ can also be attributed physical treatment of angular velocity of the string, as discussed in \cite{Durka:2019ajz,Clement:2019ghi}. Then the quantity $N$ conjugate to the angular velocity is interpreted as string angular momentum. By conducting the method of Komar integration raised in \cite{Clement:2019ghi}, alternative Smarr relation can be dervied, with ``reduced string angular momentum''. But one can prove that it can be reduced to Eq. (\ref{smax}), only by identifying the string angular velocity as the Misner potential and the string angular momentum as the Misner charge \cite{BallonBordo:2020mcs,Clement:2019ghi,BallonBordo:2019vrn}.

In above, it is obvious that we have not set $a(r_+)=0$. Correspondingly, the electromagnetic potential $A\neq 0$, and neither does the magnetic charge. If, in the other way, the regularity condition $A(r_+)=0$ is imposed on, like the Einstein case, we will have the electric first law
\begin{equation}
\delta M=T \delta S+\varphi \delta Q+\psi \delta N,
\end{equation}
together with the supplementary Smarr relation
\begin{equation}
M=2(T S+\psi N)+\varphi Q.
\end{equation}
In such situation the magnetic parameter is encoded into the electric parameter by the relation
\begin{equation}
g=-\frac{e}{2 n} \tan \left[{\rm{e}}^{-\gamma}\left(\pi-2 \arctan \frac{r_{+}}{n}\right)\right].
\end{equation}

\section{Circular motions of massive particles around the NUTty dyons}\label{sec2}
In this section, we will study the effects of the NLE dimensionless parameter $\gamma$ and the conformal scalar parameter $\alpha$ on the motion of the charged massive particle, and we will put our emphasis on the circular motion around the NUTty dyons and investigate the ISCO of the particles. Our related investigations in this section benefit from Refs. \cite{Carter:1968rr,Cebeci:2015fie,Lim:2021ejg,Lim:2020bdj}. The Lagrangian describing the motion of a charged massive particle reads
\begin{equation}
\mathcal{L}=\frac{1}{2} g_{\mu \nu} \dot{x}^{\mu} \dot{x}^{\nu}+q A_{\mu} \dot{x}^{\mu},
\end{equation}
where the overhead dot means ordinary derivative with regard to the affine parameter $\lambda$, which is connected to the proper time through the relation $\tau=\mu\lambda$, $q$ is the charge of the particle. The normalizing condition of the charged particle thus can be written as
\begin{equation}
g_{\mu \nu} \dot{x}^{\mu} \dot{x}^{\nu}=-\mu^{2},
\end{equation}
where $\mu=0,\,1$ for the massless photon and massive particle, respectively. The momenta of the charged massive particle is
\begin{equation}
P_{\mu}=\frac{\partial \mathcal{L}}{\partial \dot{x}^{\mu}}=g_{\mu \nu} \dot{x}^{\nu}+q A_{\mu},
\end{equation}
and the Hamiltonian can be obtained as
\begin{equation}
H=P_{\mu} \dot{x}^{\mu}-\mathcal{L}=\frac{1}{2} g^{\mu \nu}\left(P_{\mu}-q A_{\mu}\right)\left(P_{\nu}-q A_{\nu}\right).
\end{equation}
To solve the equation of motion for the charged massive particle, we can seek help from the Hamilton-Jacobi method, with the Hamilton-Jacobi equation being written as
\begin{equation}\label{hj1}
\frac{\partial S}{\partial \lambda}=H=\frac{1}{2} g^{\mu \nu}\left(P_{\mu}-q A_{\mu}\right)\left(P_{\nu}-q A_{\nu}\right),
\end{equation}
where $S$ is the Jacobian action which can be written in the variable-separated form as
\begin{equation}\label{hj2}
S=-\frac{1}{2} \lambda+J \phi-E t+S_{r}(r)+S_{\theta}(\theta),
\end{equation}
with $J=P_\phi\,,E=-P_t$ individually the angular momentum and the energy of the charged particle measured at the spatial infinity as constants of motion due to the symmetries of the geometry. $S_r(r)$ and $S_\theta (\theta)$ are functions of $r$ and $\theta$ to be determined. With the help of Eqs. (\ref{hj1}) and (\ref{hj2}), we can have the separated equations fulfilled by the functions $S_r$ and $S_\theta$ as
\begin{equation}
-f(r) [{\rm{d}}S_r (r)/{\rm{d}}r]^2=\frac{K}{n^2+r^2}-\frac{[q a(r)+E]^2}{f(r)}+\mu^2,
\end{equation}
\begin{equation}
[{\rm{d}}S_\theta (\theta)/{\rm{d}}r]^2=K-\left[J \csc (\theta )+2 E n \cot (\theta )\right]^2,
\end{equation}
where $K$ is a separation constant.

\begin{figure*}[htpb!]
\begin{center}
\includegraphics[width=3.4in,angle=0]{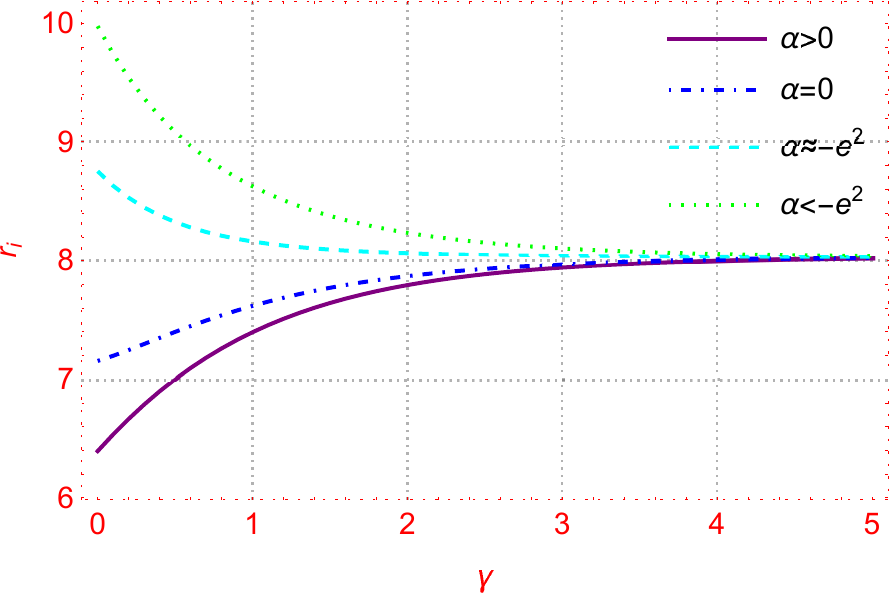}
\includegraphics[width=3.5in,angle=0]{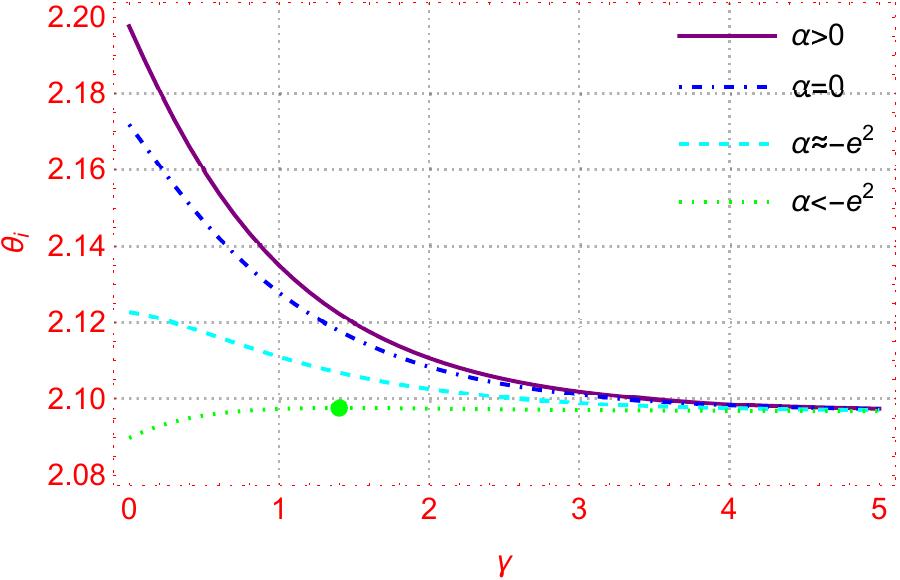}
\end{center}
\vspace{-5mm}
 \caption {Variations of the radius $r_i$ and the latitude $\theta_i$ for the massive particle on the ISCO circular orbits with respect to the NLE parameter $\gamma$ for $m=1\,,n=1,\,e=1/2\,,g=1/2,\,q=1/2$. The solid purple,  dot-and-dash blue, dashed cyan, and dotted green lines are for the $\alpha=0.1>0$, $\alpha=0$, $\alpha=-0.26\lessapprox -e^2$ and $\alpha=-0.5<-e^2$ cases, respectively. The green dot denotes the extreme point on the green curve.}\label{pic1}
\end{figure*}

According to the relations
\begin{equation}
P_{r}=\frac{\partial S}{\partial r}, P_{\theta}=\frac{\partial S}{\partial \theta},
\end{equation}
we further have
\begin{equation}
P_r=f(r)^{-1}\sqrt{R(r)},
\end{equation}
\begin{equation}
P_\theta=\sqrt{\Theta(\theta)},
\end{equation}
where we have denoted
\begin{equation}\label{ep1}
R(r)=[E+q a(r)]^2-\left(\frac{K}{n^2+r^2}+\mu^2\right)f(r),
\end{equation}
\begin{equation}\label{ep2}
\Theta(\theta)=K-(2 n E \cot\theta+J\sin^{-1}\theta)^2,
\end{equation}
which are radial and latitudinal effective potentials of the particles. As a result, the Jacobian action as a solution of the Hamilton-Jacobi equation can be written in the form
\begin{equation}
\begin{aligned}
S=-\frac{1}{2} \mu^{2} \tau-E t+J \phi +\int^{\theta}\sqrt{\Theta(\theta)} {\rm{d}} \theta+\int^{r} \frac{\sqrt{R(r)}}{f(r)} {\rm{d}} r.
\end{aligned}
\end{equation}
After differentiating the Jacobian action relative to the constants of the motion $K,\,\mu,\,E\,,J$ for the particle, we can obtain the integrated forms of the geodesics for the particle, expressed as
\begin{equation}
\int^{\theta} \frac{{\rm{d}} \theta}{\sqrt{\Theta}}=\int^{r} \frac{{\rm{d}} r}{(n^2+r^2)\sqrt{R}},
\end{equation}
\begin{equation}
\tau=\int^{r} \frac{{\rm{d}} r}{\sqrt{R}},
\end{equation}
\begin{equation}
\begin{aligned}
t=&\int^{\theta} \frac{-2n\cot\theta\left(2 n E\cot\theta+J\sin^{-1}\theta\right) {\rm{d}} \theta}{\sqrt{\Theta(\theta)}}\\&+\int^r \frac{E+q a(r)}{f(r)\sqrt{R(r)}}{\rm{d}}r,
\end{aligned}
\end{equation}
\begin{equation}
\phi=\int^\theta \frac{2nE\cot\theta+J\sin^{-1}\theta}{\sqrt{\Theta(\theta)}\sin\theta}{\rm{d}}\theta.
\end{equation}
Their first-order forms can be explicitly got as
\begin{equation}
\frac{{\rm{d}}t}{{\rm{d}}\tau}=\frac{-2n\cot\theta(2nE\cot\theta+J\sin^{-1}\theta)}{n^2+r^2}+\frac{E+qa(r)}{f(r)},
\end{equation}
\begin{equation}
\frac{{\rm{d}}r}{{\rm{d}}\tau}=\sqrt{R(r)},
\end{equation}
\begin{equation}
\frac{{\rm{d}}\theta}{{\rm{d}}\tau}=\frac{\sqrt{\Theta(\theta)}}{n^2+r^2},
\end{equation}
\begin{equation}
\frac{{\rm{d}}\phi}{{\rm{d}}\tau}=\frac{2nE\cot\theta+J\sin^{-1}\theta}{(n^2+r^2)\sin\theta}.
\end{equation}

With these equations of motion in hand, we can immediately find the locations of the ISCO for the charged massive particles around the NUTty dyons. To that end, we should let
\begin{equation}
R\left(r_{i}\right)=0, \quad \left.\frac{{\rm{d}}R\left(r\right)}{{\rm{d}}r}\right|_{r=r_i}=0, \quad \left.\frac{{\rm{d}}^2 R\left(r\right)}{{\rm{d}}r^2}\right|_{r=r_i}=0.
\end{equation}
The first one is satisfied by the radial turning point; the second one together with the first one produces the circular orbit with constant radius; the last one restricts that the circular orbit is marginally stable, or in other words, it provides the innermost circular orbit. Besides, the latitudinal conditions
\begin{equation}
\Theta\left(\theta_{i}\right)=0, \quad \left.\frac{{\rm{d}}\Theta\left(\theta\right)}{{\rm{d}}\theta}\right|_{\theta=\theta_i}=0, \quad \left.\frac{{\rm{d}}^2 \Theta\left(\theta\right)}{{\rm{d}}\theta^2}\right|_{\theta=\theta_i}=0
\end{equation}
should also be satisfied, which ensure that the particle is stably located on the position with constant latitude. We numerically calculate the ISCO for the massive particles and obtain the related parameters, of which two representative ones, the radius and the latitude, are shown in Fig. \ref{pic1}. When the NLE parameter is large enough, all parameters tend to be constant. This is easy to understand once we glimpse at the blackening factor Eq. (\ref{blf}) where $\gamma$ penetrates. Besides, from the diagrams, we can also find other important properties. First, the changing tendency of the ISCO radius depends on the value range of the conformal scalar parameter $\alpha$. That is, if $\alpha\geqslant 0$, the ISCO radius increases with respect to the increasing NLE parameter $\gamma$; otherwise, if $\alpha$ belongs to the other branch where $\alpha<-e^2$, the ISCO radius behaves contrarily. Secondly, due to the emergence of the NUT parameter $n$, the ISCO will not locate on the equatorial plane. Here we see that how the conformal scalar parameter and the NLE parameter interplay to change the circular plane. With a non-negative conformal scalar parameter, the latitude of the ISCO plane decreases monotonically with respect to the increasing NLE parameter, and this style can also be shared by the case where $\alpha<-e^2$. However, if $\alpha$ is small enough, the effect of the NLE parameter $\gamma$ changes. There will be an extreme point where the effect of the NLE parameter $\gamma$ dominates. Lastly, when the NLE parameter is kept unchanged and the conformal scalar parameter increases, the ISCO radius increases while the ISCO latitude decreases. In other words, the closer the ISCO to the equatorial plane, the larger the ISCO radius.

\section{Shadows of the NUTty dyons}\label{sec3}
In this section, we will explore the effects of the NLE dimensionless parameter $\gamma$ and the conformal scalar parameter $\alpha$ on the shadow of the conformally scalar NUTty dyons. Refs. \cite{Cunha:2016bpi,Konoplya:2019sns,Grenzebach:2014fha,Zhang:2020xub,Grenzebach:2015oea,Perlick:2021aok,Wei:2018xks,Li:2020drn} are important literature to advance our work here. For simplicity, we set $m=1, E=1$ in what follows. Notice that $\mu=0$ for the photon, whose radial and latitude effective potentials are described by
Eqs. (\ref{ep1}) and (\ref{ep2}) individually. Firstly we should obtain the circular orbits of the photons around the NUTty dyon black hole, which demands
\begin{equation}
R\left(r_{p}\right)=0, \quad \left.\frac{{\rm{d}}R\left(r\right)}{{\rm{d}}r}\right|_{r=r_p}=0, \quad \left.\frac{{\rm{d}}^2 R\left(r\right)}{{\rm{d}}r^2}\right|_{r=r_p}>0,
\end{equation}
where the last condition means that the circular orbit of the photon is radically unstable. We should also have
\begin{equation}
\Theta\left(\theta_{p}\right)=0, \quad \left.\frac{{\rm{d}}\Theta\left(\theta\right)}{{\rm{d}}\theta}\right|_{\theta=\theta_p}=0, \quad \left.\frac{{\rm{d}}^2 \Theta\left(\theta\right)}{{\rm{d}}\theta^2}\right|_{\theta=\theta_p}<0,
\end{equation}
where the last condition denotes that the orbit is latitudinally stable. Using them, we get the related characterized parameters of the photons on the circular orbit as
\begin{equation}
K_p=4 n^{2} \tan\theta_p^{2},
\end{equation}
\begin{equation}
J_p=-2 n \sec\theta_p,
\end{equation}
\begin{equation}
\tan\theta_p=\frac{\sqrt{n^2+r_p^2}}{2 n \sqrt{f(r_p)}}.
\end{equation}
Besides, the radius of the photon is restricted by the equation
\begin{equation}
\begin{aligned}
-f'(r_p) +\frac{2 r_p f(r_p)}{n^2+r_p^2}=0.
\end{aligned}
\end{equation}

\begin{figure*}[htpb!]
\begin{center}
\includegraphics[width=4in,angle=0]{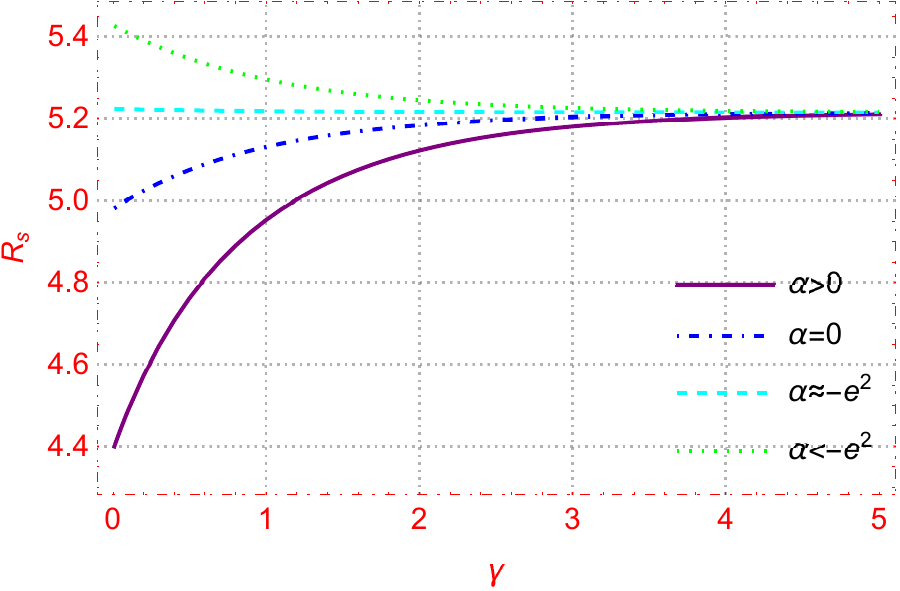}
\end{center}
\vspace{-5mm}
 \caption {Variations of the shadow radius with respect to the NLE parameter $\gamma$ for $m=1\,,n=1/10,\,e=1/2,\,g=1/2$. The solid purple,  dot-and-dash blue, dashed cyan, and dotted green lines are for the $\alpha=0.5>0$, $\alpha=0$, $\alpha=-0.26\lessapprox -e^2$ and $\alpha=-0.5<-e^2$ cases, respectively. Note that the cyan curve is not horizontal; it shows that $R_s$ decreases  slowly with increasing $\gamma$.}\label{pic2}
\end{figure*}

The basis $\left\{\hat{e}_{(t)}, \hat{e}_{(r)}, \hat{e}_{(\theta)}, \hat{e}_{(\varphi)}\right\}$ for the observer can be projected onto basis $\left\{\partial_{t}, \partial_{r}, \partial_{\theta}, \partial_{\varphi}\right\}$ for the spacetime. One usually used orthogonal and normalized tetrad for the observer reads \cite{Cunha:2016bpi}
\begin{equation}
\begin{aligned}
\hat{e}_{(t)} &=\sqrt{\frac{g_{\phi \phi}}{g_{t \phi}^{2}-g_{t t} g_{\phi \phi}}}\left(\partial_{t}-\frac{g_{t \phi}}{g_{\phi \phi}} \partial_{\phi}\right), \\
\hat{e}_{(r)} &=\frac{1}{\sqrt{g_{r r}}} \partial_{r}, \\
\hat{e}_{(\theta)} &=\frac{1}{\sqrt{g_{\theta \theta}}} \partial_{\theta}, \\
\hat{e}_{(\phi)} &=\frac{1}{\sqrt{g_{\phi \phi}}} \partial_{\phi},
\end{aligned}
\end{equation}
which corresponds to a zero-angular-momentum-observer (ZAMO). An observer in this frame moves with an angular velocity $-g_{t \phi} / g_{\phi \phi}$ relative to spatial infinity, due to the dragging effect of the black hole.

The locally measured four-momentum of the photon can be obtained by its projecting onto $\hat{e}_{(t)}^{\mu}$,
\begin{equation}
\begin{aligned}
&p^{(t)}=-p_{\mu} \hat{e}_{(t)}^{\mu}, \\
&p^{(i)}=p_{\mu} \hat{e}_{(i)}^{\mu},
\end{aligned}
\end{equation}
with $i=r, \theta, \phi$.

For the massless photon, we have 
\begin{equation}
\left[p^{(t)}\right]^{2}=\left[p^{(r)}\right]^{2}+\left[p^{(\theta)}\right]^{2}+\left[p^{(\varphi)}\right]^{2}.
\end{equation}
So the observation angles can be defined as
\begin{equation}
\begin{aligned}
&p^{(r)}=p^{(t)} \cos \tilde{\alpha} \cos \beta, \\
&p^{(\theta)}=p^{(t)} \sin \tilde{\alpha}, \\
&p^{(\phi)}=p^{(t)} \cos \tilde{\alpha} \sin \beta,
\end{aligned}
\end{equation}
 Explicitly, the angular coordinates can be written as
\begin{equation}
\sin \tilde{\alpha}=\frac{p^{(\theta)}}{p^{(t)}},
\end{equation}
\begin{equation}
\tan \beta=\frac{p^{(\phi)}}{p^{(r)}}.
\end{equation}

The perimeter radius of a circumference at constant $\theta$ and $r$ can be defined by
\begin{equation}
\tilde{r} \equiv \frac{1}{2 \pi}\int_{0}^{2 \pi} \sqrt{g_{\phi \phi}} {\rm{d}} \phi=\sqrt{g_{\phi \phi}}.
\end{equation}
Then the Cartesian coordinate on the sky plane of the observer can be written as
\begin{equation}
\begin{aligned}
x &\equiv-\tilde{r} \beta=-\tilde{r}\arctan \left[\frac{p^{(\phi)}}{p^{(r)}}\right]\\&=\left.\sqrt{g_{\phi \phi}}\arctan\frac{f(r)\sqrt{g_{rr}}}{\sqrt{R(r)g_{\phi\phi}}}\right|_{(r_o\,,\theta_o)},
\end{aligned}
\end{equation}
\begin{equation}
\begin{aligned}
y &\equiv \tilde{r} \tilde{\alpha}=\tilde{r}\arctan \left[\frac{p^{(\theta)}}{p^{(r)}}\right]\\&=\sqrt{\left(n^2+r^2\right)\sin ^2\theta -4 n^2 f(r) \cos ^2\theta} \\&\left.\quad\times \arcsin \frac{\sqrt{\Theta (\theta )}}{(\zeta -\iota J) \sqrt{n^2+r^2}}\right|_{(r_o\,,\theta_o)},
\end{aligned}
\end{equation}
where $r_o$ and $\theta_o$ are individually radial coordinate and inclination angle of the observer. Here we have denoted $\zeta \equiv \hat{e}_{(t)}^{t}, \iota \equiv \hat{e}_{(t)}^{\phi}$, which are evaluated at the photon orbit.

At very large distance, we have
\begin{equation}
\lim_{r_o\to \infty}x \equiv X=-J,
\end{equation}
\begin{equation}
\lim_{r_o\to \infty}y\equiv Y=\sin\theta_o\sqrt{\Theta(\theta_o)}.
\end{equation}
To directly reflect the effects of the conformal scalar parameter and the NLE parameter on the shadow of the black hole, we here set the observer on the plane determined by the photon doing circular motion, so that $Y=0$. Then we have the shadow radius of the black hole
\begin{equation}\begin{aligned}
R_s=|X|=|2 n \sec\theta_p|=2 n\sqrt{1+\frac{n^2+r_p^2 }{2 n^2 f(r_p)}}.
\end{aligned}\end{equation}
One can check that when $n=\gamma=g=\alpha=0$, $R_s$ reduces to $3\sqrt{3}$, which is the shadow radius of the Schwarzschild black hole. To visually see the effect of the interplay between the NLE parameter and the conformal scalar parameter on the shadow radius of the black hole, we plot Fig. \ref{pic2}, from which we can see that: (1) if $\alpha\geqslant 0$, the shadow radius increases with the increasing NLE parameter, but the changing style becomes opposite if $\alpha<-e^2$; (2) the shadow radius decreases if the conformal parameter $\alpha$ increases.

\section{Conclusions}
In this paper, we first found a NUTty dyon black hole with NUT charge, electric and magnetic charges, as well as the conformal scalar parameter and the NLE parameter. This is a nontrivial conformally scalar black hole solution in the conformal electrodynamics, incorporating the characteristics of the conformally scalar coupled gravity and the conformal invariance and $SO(2)$ electromagnetic duality of the ModMax theory. The Euclidean method was used to calculate the Gibbs free energy of the black hole and the cohomogeneity thermodynamics for the asymptotically flat black hole was formulated. The conformal scalar incorporates into the mass and entropy of the black hole but does not enter the first law of the black hole as an independent variable, though it is a primary hair as it does exist independently even with vanishing electric charge. 

To visualize the strong gravitational effects of the conformal scalar hair and the NLE and their interplay, we further studied the ISCO of the charged massive particle around the black hole as well as the shadow formed by the photons. We calculated the characteristic radius and latitude of the ISCO for the charged massive particle. We showed that if the black hole is Reissner-Nordström-like, corresponding to a non-negative conformal scalar parameter, the radius increases but the latitude decreases with respect to the increasing NLE parameter. If the black hole is Schwarzschild-like, endowed with a negative scalar hair, the radius of the ISCO decreases with the increasing NLE parameter, but the latitude of the ISCO may increase first and then decrease. This is due to that the nonlinearity of the electromagnetic field counterbalance the effect of the scalar hair. Other than that, we found that the greater the conformal scalar parameter, the larger the ISCO radius and the nearer the circular plane apart from the equatorial plane.

Choosing the ZAMO, we obtained the shadow radius of the NUTty dyon black hole. On the one hand, not like most other black holes whose unstable circular photon orbit locates on the equatorial plane, the circular orbit of the photons around the NUTty dyon black hole derivates from the equator; on the other hand, $-g_{t \phi} / g_{\phi \phi}\neq 0$ for the black hole, so we calculate its shadow with a method similar to the one for the Kerr black hole, albeit the black hole has vanishing angular momentum, resulting in that the formulae of its shadow radius being similar to the Schwarzschild one. This is quite unexpected! When the nonlinearity of the conformal electrodynamics increases, the shadow radius of the NUTty dyon with positive conformal scalar parameter also increases, but the one with nonpositive parameter decreases. Besides, the greater the conformal scalar parameter, the larger the shadow.

In summary, we found a NUTty dyon solution with conformal scalar hair in conformal electrodynamics and showed the strong gravitational effects of the interplay between the conformal scalar hair and the nonlinear electrodynamics. It is worthwhile to further explore the gravitational effects of the novel conformal electrodynamics in other theories beyond GR and our work here may be helpful.

\section*{Acknowledgements}
M. Z. is supported by the National Natural Science Foundation of China (Grant No. 12005080) and Young Talents Foundation of Jiangxi Normal University (Grant No. 12020779).  J. J.  is supported by the National Natural Science Foundation of China (Grants No. 11775022 and No. 11873044).


\begin{thebibliography}{49}%
\makeatletter
\providecommand \@ifxundefined [1]{%
 \@ifx{#1\undefined}
}%
\providecommand \@ifnum [1]{%
 \ifnum #1\expandafter \@firstoftwo
 \else \expandafter \@secondoftwo
 \fi
}%
\providecommand \@ifx [1]{%
 \ifx #1\expandafter \@firstoftwo
 \else \expandafter \@secondoftwo
 \fi
}%
\providecommand \natexlab [1]{#1}%
\providecommand \enquote  [1]{``#1''}%
\providecommand \bibnamefont  [1]{#1}%
\providecommand \bibfnamefont [1]{#1}%
\providecommand \citenamefont [1]{#1}%
\providecommand \href@noop [0]{\@secondoftwo}%
\providecommand \href [0]{\begingroup \@sanitize@url \@href}%
\providecommand \@href[1]{\@@startlink{#1}\@@href}%
\providecommand \@@href[1]{\endgroup#1\@@endlink}%
\providecommand \@sanitize@url [0]{\catcode `\\12\catcode `\$12\catcode
  `\&12\catcode `\#12\catcode `\^12\catcode `\_12\catcode `\%12\relax}%
\providecommand \@@startlink[1]{}%
\providecommand \@@endlink[0]{}%
\providecommand \url  [0]{\begingroup\@sanitize@url \@url }%
\providecommand \@url [1]{\endgroup\@href {#1}{\urlprefix }}%
\providecommand \urlprefix  [0]{URL }%
\providecommand \Eprint [0]{\href }%
\providecommand \doibase [0]{https://doi.org/}%
\providecommand \selectlanguage [0]{\@gobble}%
\providecommand \bibinfo  [0]{\@secondoftwo}%
\providecommand \bibfield  [0]{\@secondoftwo}%
\providecommand \translation [1]{[#1]}%
\providecommand \BibitemOpen [0]{}%
\providecommand \bibitemStop [0]{}%
\providecommand \bibitemNoStop [0]{.\EOS\space}%
\providecommand \EOS [0]{\spacefactor3000\relax}%
\providecommand \BibitemShut  [1]{\csname bibitem#1\endcsname}%
\let\auto@bib@innerbib\@empty
%</preamble>
\bibitem [{\citenamefont {Aaboud}\ \emph {et~al.}(2017)\citenamefont {Aaboud}
  \emph {et~al.}}]{ATLAS:2017fur}%
  \BibitemOpen
  \bibfield  {author} {\bibinfo {author} {\bibfnamefont {M.}~\bibnamefont
  {Aaboud}} \emph {et~al.} (\bibinfo {collaboration} {ATLAS}),\ }\bibfield
  {title} {\bibinfo {title} {{Evidence for light-by-light scattering in
  heavy-ion collisions with the ATLAS detector at the LHC}},\ }\href
  {https://doi.org/10.1038/nphys4208} {\bibfield  {journal} {\bibinfo
  {journal} {Nature Phys.}\ }\textbf {\bibinfo {volume} {13}},\ \bibinfo
  {pages} {852} (\bibinfo {year} {2017})},\ \Eprint
  {https://arxiv.org/abs/1702.01625} {arXiv:1702.01625 [hep-ex]} \BibitemShut
  {NoStop}%
\bibitem [{\citenamefont {Born}(1934)}]{Born:1934ji}%
  \BibitemOpen
  \bibfield  {author} {\bibinfo {author} {\bibfnamefont {M.}~\bibnamefont
  {Born}},\ }\bibfield  {title} {\bibinfo {title} {{Quantum theory of the
  electromagnetic field}},\ }\href {https://doi.org/10.1098/rspa.1934.0010}
  {\bibfield  {journal} {\bibinfo  {journal} {Proc. Roy. Soc. Lond. A}\
  }\textbf {\bibinfo {volume} {143}},\ \bibinfo {pages} {410} (\bibinfo {year}
  {1934})}\BibitemShut {NoStop}%
\bibitem [{\citenamefont {Born}\ and\ \citenamefont
  {Infeld}(1934)}]{Born:1934gh}%
  \BibitemOpen
  \bibfield  {author} {\bibinfo {author} {\bibfnamefont {M.}~\bibnamefont
  {Born}}\ and\ \bibinfo {author} {\bibfnamefont {L.}~\bibnamefont {Infeld}},\
  }\bibfield  {title} {\bibinfo {title} {{Foundations of the new field
  theory}},\ }\href {https://doi.org/10.1098/rspa.1934.0059} {\bibfield
  {journal} {\bibinfo  {journal} {Proc. Roy. Soc. Lond. A}\ }\textbf {\bibinfo
  {volume} {144}},\ \bibinfo {pages} {425} (\bibinfo {year}
  {1934})}\BibitemShut {NoStop}%
\bibitem [{\citenamefont {Heisenberg}\ and\ \citenamefont
  {Euler}(1936)}]{Heisenberg:1936nmg}%
  \BibitemOpen
  \bibfield  {author} {\bibinfo {author} {\bibfnamefont {W.}~\bibnamefont
  {Heisenberg}}\ and\ \bibinfo {author} {\bibfnamefont {H.}~\bibnamefont
  {Euler}},\ }\bibfield  {title} {\bibinfo {title} {{Consequences of Dirac's
  theory of positrons}},\ }\href {https://doi.org/10.1007/BF01343663}
  {\bibfield  {journal} {\bibinfo  {journal} {Z. Phys.}\ }\textbf {\bibinfo
  {volume} {98}},\ \bibinfo {pages} {714} (\bibinfo {year} {1936})},\ \Eprint
  {https://arxiv.org/abs/physics/0605038} {arXiv:physics/0605038} \BibitemShut
  {NoStop}%
\bibitem [{\citenamefont {Bandos}\ \emph {et~al.}(2020)\citenamefont {Bandos},
  \citenamefont {Lechner}, \citenamefont {Sorokin},\ and\ \citenamefont
  {Townsend}}]{Bandos:2020jsw}%
  \BibitemOpen
  \bibfield  {author} {\bibinfo {author} {\bibfnamefont {I.}~\bibnamefont
  {Bandos}}, \bibinfo {author} {\bibfnamefont {K.}~\bibnamefont {Lechner}},
  \bibinfo {author} {\bibfnamefont {D.}~\bibnamefont {Sorokin}},\ and\ \bibinfo
  {author} {\bibfnamefont {P.~K.}\ \bibnamefont {Townsend}},\ }\bibfield
  {title} {\bibinfo {title} {{A non-linear duality-invariant conformal
  extension of Maxwell's equations}},\ }\href
  {https://doi.org/10.1103/PhysRevD.102.121703} {\bibfield  {journal} {\bibinfo
   {journal} {Phys. Rev. D}\ }\textbf {\bibinfo {volume} {102}},\ \bibinfo
  {pages} {121703} (\bibinfo {year} {2020})},\ \Eprint
  {https://arxiv.org/abs/2007.09092} {arXiv:2007.09092 [hep-th]} \BibitemShut
  {NoStop}%
\bibitem [{\citenamefont {Kosyakov}(2020)}]{Kosyakov:2020wxv}%
  \BibitemOpen
  \bibfield  {author} {\bibinfo {author} {\bibfnamefont {B.~P.}\ \bibnamefont
  {Kosyakov}},\ }\bibfield  {title} {\bibinfo {title} {{Nonlinear
  electrodynamics with the maximum allowable symmetries}},\ }\href
  {https://doi.org/10.1016/j.physletb.2020.135840} {\bibfield  {journal}
  {\bibinfo  {journal} {Phys. Lett. B}\ }\textbf {\bibinfo {volume} {810}},\
  \bibinfo {pages} {135840} (\bibinfo {year} {2020})},\ \Eprint
  {https://arxiv.org/abs/2007.13878} {arXiv:2007.13878 [hep-th]} \BibitemShut
  {NoStop}%
\bibitem [{\citenamefont {Ballon~Bordo}\ \emph {et~al.}(2021)\citenamefont
  {Ballon~Bordo}, \citenamefont {Kubiz\v{n}\'ak},\ and\ \citenamefont
  {Perche}}]{BallonBordo:2020jtw}%
  \BibitemOpen
  \bibfield  {author} {\bibinfo {author} {\bibfnamefont {A.}~\bibnamefont
  {Ballon~Bordo}}, \bibinfo {author} {\bibfnamefont {D.}~\bibnamefont
  {Kubiz\v{n}\'ak}},\ and\ \bibinfo {author} {\bibfnamefont {T.~R.}\
  \bibnamefont {Perche}},\ }\bibfield  {title} {\bibinfo {title} {{Taub-NUT
  solutions in conformal electrodynamics}},\ }\href
  {https://doi.org/10.1016/j.physletb.2021.136312} {\bibfield  {journal}
  {\bibinfo  {journal} {Phys. Lett. B}\ }\textbf {\bibinfo {volume} {817}},\
  \bibinfo {pages} {136312} (\bibinfo {year} {2021})},\ \Eprint
  {https://arxiv.org/abs/2011.13398} {arXiv:2011.13398 [hep-th]} \BibitemShut
  {NoStop}%
\bibitem [{\citenamefont {Flores-Alfonso}\ \emph {et~al.}(2021)\citenamefont
  {Flores-Alfonso}, \citenamefont {Gonz\'alez-Morales}, \citenamefont
  {Linares},\ and\ \citenamefont {Maceda}}]{Flores-Alfonso:2020euz}%
  \BibitemOpen
  \bibfield  {author} {\bibinfo {author} {\bibfnamefont {D.}~\bibnamefont
  {Flores-Alfonso}}, \bibinfo {author} {\bibfnamefont {B.~A.}\ \bibnamefont
  {Gonz\'alez-Morales}}, \bibinfo {author} {\bibfnamefont {R.}~\bibnamefont
  {Linares}},\ and\ \bibinfo {author} {\bibfnamefont {M.}~\bibnamefont
  {Maceda}},\ }\bibfield  {title} {\bibinfo {title} {{Black holes and
  gravitational waves sourced by non-linear duality rotation-invariant
  conformal electromagnetic matter}},\ }\href
  {https://doi.org/10.1016/j.physletb.2020.136011} {\bibfield  {journal}
  {\bibinfo  {journal} {Phys. Lett. B}\ }\textbf {\bibinfo {volume} {812}},\
  \bibinfo {pages} {136011} (\bibinfo {year} {2021})},\ \Eprint
  {https://arxiv.org/abs/2011.10836} {arXiv:2011.10836 [gr-qc]} \BibitemShut
  {NoStop}%
\bibitem [{\citenamefont {Bokuli\'c}\ \emph {et~al.}(2021)\citenamefont
  {Bokuli\'c}, \citenamefont {Juri\'c},\ and\ \citenamefont
  {Smoli\'c}}]{Bokulic:2021dtz}%
  \BibitemOpen
  \bibfield  {author} {\bibinfo {author} {\bibfnamefont {A.}~\bibnamefont
  {Bokuli\'c}}, \bibinfo {author} {\bibfnamefont {T.}~\bibnamefont {Juri\'c}},\
  and\ \bibinfo {author} {\bibfnamefont {I.}~\bibnamefont {Smoli\'c}},\
  }\bibfield  {title} {\bibinfo {title} {{Black hole thermodynamics in the
  presence of nonlinear electromagnetic fields}},\ }\href@noop {} {\  (\bibinfo
  {year} {2021})},\ \Eprint {https://arxiv.org/abs/2102.06213}
  {arXiv:2102.06213 [gr-qc]} \BibitemShut {NoStop}%
\bibitem [{\citenamefont {Ruffini}\ and\ \citenamefont
  {Wheeler}(1971)}]{Ruffini:1971bza}%
  \BibitemOpen
  \bibfield  {author} {\bibinfo {author} {\bibfnamefont {R.}~\bibnamefont
  {Ruffini}}\ and\ \bibinfo {author} {\bibfnamefont {J.~A.}\ \bibnamefont
  {Wheeler}},\ }\bibfield  {title} {\bibinfo {title} {{Introducing the black
  hole}},\ }\href {https://doi.org/10.1063/1.3022513} {\bibfield  {journal}
  {\bibinfo  {journal} {Phys. Today}\ }\textbf {\bibinfo {volume} {24}},\
  \bibinfo {pages} {30} (\bibinfo {year} {1971})}\BibitemShut {NoStop}%
\bibitem [{\citenamefont {Herdeiro}\ and\ \citenamefont
  {Radu}(2015)}]{Herdeiro:2015waa}%
  \BibitemOpen
  \bibfield  {author} {\bibinfo {author} {\bibfnamefont {C.~A.~R.}\
  \bibnamefont {Herdeiro}}\ and\ \bibinfo {author} {\bibfnamefont
  {E.}~\bibnamefont {Radu}},\ }\bibfield  {title} {\bibinfo {title}
  {{Asymptotically flat black holes with scalar hair: a review}},\ }\href
  {https://doi.org/10.1142/S0218271815420146} {\bibfield  {journal} {\bibinfo
  {journal} {Int. J. Mod. Phys. D}\ }\textbf {\bibinfo {volume} {24}},\
  \bibinfo {pages} {1542014} (\bibinfo {year} {2015})},\ \Eprint
  {https://arxiv.org/abs/1504.08209} {arXiv:1504.08209 [gr-qc]} \BibitemShut
  {NoStop}%
\bibitem [{\citenamefont {Bocharova}\ \emph {et~al.}(1970)\citenamefont
  {Bocharova}, \citenamefont {Bronnikov},\ and\ \citenamefont
  {Melnikov}}]{bocharova1970exact}%
  \BibitemOpen
  \bibfield  {author} {\bibinfo {author} {\bibfnamefont {N.}~\bibnamefont
  {Bocharova}}, \bibinfo {author} {\bibfnamefont {K.}~\bibnamefont
  {Bronnikov}},\ and\ \bibinfo {author} {\bibfnamefont {V.}~\bibnamefont
  {Melnikov}},\ }\bibfield  {title} {\bibinfo {title} {An exact solution of the
  system of einstein equations and mass-free scalar field, vestn},\ }\href@noop
  {} {\bibfield  {journal} {\bibinfo  {journal} {Mosk. univ. Fiz. astron}\
  }\textbf {\bibinfo {volume} {6}},\ \bibinfo {pages} {706} (\bibinfo {year}
  {1970})}\BibitemShut {NoStop}%
\bibitem [{\citenamefont {Bekenstein}(1975)}]{bekenstein1975black}%
  \BibitemOpen
  \bibfield  {author} {\bibinfo {author} {\bibfnamefont {J.~D.}\ \bibnamefont
  {Bekenstein}},\ }\bibfield  {title} {\bibinfo {title} {Black holes with
  scalar charge},\ }\href@noop {} {\bibfield  {journal} {\bibinfo  {journal}
  {Annals of Physics}\ }\textbf {\bibinfo {volume} {91}},\ \bibinfo {pages}
  {75} (\bibinfo {year} {1975})}\BibitemShut {NoStop}%
\bibitem [{\citenamefont {Anabalon}\ and\ \citenamefont
  {Cisterna}(2012)}]{Anabalon:2012tu}%
  \BibitemOpen
  \bibfield  {author} {\bibinfo {author} {\bibfnamefont {A.}~\bibnamefont
  {Anabalon}}\ and\ \bibinfo {author} {\bibfnamefont {A.}~\bibnamefont
  {Cisterna}},\ }\bibfield  {title} {\bibinfo {title} {{Asymptotically (anti)
  de Sitter Black Holes and Wormholes with a Self Interacting Scalar Field in
  Four Dimensions}},\ }\href {https://doi.org/10.1103/PhysRevD.85.084035}
  {\bibfield  {journal} {\bibinfo  {journal} {Phys. Rev. D}\ }\textbf {\bibinfo
  {volume} {85}},\ \bibinfo {pages} {084035} (\bibinfo {year} {2012})},\
  \Eprint {https://arxiv.org/abs/1201.2008} {arXiv:1201.2008 [hep-th]}
  \BibitemShut {NoStop}%
\bibitem [{\citenamefont {Simovic}\ \emph {et~al.}(2020)\citenamefont
  {Simovic}, \citenamefont {Fusco},\ and\ \citenamefont
  {Mann}}]{Simovic:2020dke}%
  \BibitemOpen
  \bibfield  {author} {\bibinfo {author} {\bibfnamefont {F.}~\bibnamefont
  {Simovic}}, \bibinfo {author} {\bibfnamefont {D.}~\bibnamefont {Fusco}},\
  and\ \bibinfo {author} {\bibfnamefont {R.~B.}\ \bibnamefont {Mann}},\
  }\bibfield  {title} {\bibinfo {title} {{Thermodynamics of de Sitter Black
  Holes with Conformally Coupled Scalar Fields}}\ }\href
  {https://doi.org/10.1007/JHEP02(2021)219} {10.1007/JHEP02(2021)219} (\bibinfo
  {year} {2020}),\ \Eprint {https://arxiv.org/abs/2008.07593} {arXiv:2008.07593
  [gr-qc]} \BibitemShut {NoStop}%
\bibitem [{\citenamefont {Zou}\ and\ \citenamefont
  {Myung}(2020{\natexlab{a}})}]{Zou:2020zxq}%
  \BibitemOpen
  \bibfield  {author} {\bibinfo {author} {\bibfnamefont {D.-C.}\ \bibnamefont
  {Zou}}\ and\ \bibinfo {author} {\bibfnamefont {Y.~S.}\ \bibnamefont
  {Myung}},\ }\bibfield  {title} {\bibinfo {title} {{Radial perturbations of
  the scalarized black holes in Einstein-Maxwell-conformally coupled scalar
  theory}},\ }\href {https://doi.org/10.1103/PhysRevD.102.064011} {\bibfield
  {journal} {\bibinfo  {journal} {Phys. Rev. D}\ }\textbf {\bibinfo {volume}
  {102}},\ \bibinfo {pages} {064011} (\bibinfo {year} {2020}{\natexlab{a}})},\
  \Eprint {https://arxiv.org/abs/2005.06677} {arXiv:2005.06677 [gr-qc]}
  \BibitemShut {NoStop}%
\bibitem [{\citenamefont {Martinez}\ and\ \citenamefont
  {Zanelli}(1996)}]{Martinez:1996gn}%
  \BibitemOpen
  \bibfield  {author} {\bibinfo {author} {\bibfnamefont {C.}~\bibnamefont
  {Martinez}}\ and\ \bibinfo {author} {\bibfnamefont {J.}~\bibnamefont
  {Zanelli}},\ }\bibfield  {title} {\bibinfo {title} {{Conformally dressed
  black hole in (2+1)-dimensions}},\ }\href
  {https://doi.org/10.1103/PhysRevD.54.3830} {\bibfield  {journal} {\bibinfo
  {journal} {Phys. Rev. D}\ }\textbf {\bibinfo {volume} {54}},\ \bibinfo
  {pages} {3830} (\bibinfo {year} {1996})},\ \Eprint
  {https://arxiv.org/abs/gr-qc/9604021} {arXiv:gr-qc/9604021} \BibitemShut
  {NoStop}%
\bibitem [{\citenamefont {Astorino}(2013)}]{Astorino:2013sfa}%
  \BibitemOpen
  \bibfield  {author} {\bibinfo {author} {\bibfnamefont {M.}~\bibnamefont
  {Astorino}},\ }\bibfield  {title} {\bibinfo {title} {{C-metric with a
  conformally coupled scalar field in a magnetic universe}},\ }\href
  {https://doi.org/10.1103/PhysRevD.88.104027} {\bibfield  {journal} {\bibinfo
  {journal} {Phys. Rev. D}\ }\textbf {\bibinfo {volume} {88}},\ \bibinfo
  {pages} {104027} (\bibinfo {year} {2013})},\ \Eprint
  {https://arxiv.org/abs/1307.4021} {arXiv:1307.4021 [gr-qc]} \BibitemShut
  {NoStop}%
\bibitem [{\citenamefont {Chowdhury}(2019)}]{Chowdhury:2019uwi}%
  \BibitemOpen
  \bibfield  {author} {\bibinfo {author} {\bibfnamefont {A.}~\bibnamefont
  {Chowdhury}},\ }\bibfield  {title} {\bibinfo {title} {{Hawking emission of
  charged particles from an electrically charged spherical black hole with
  scalar hair}},\ }\href {https://doi.org/10.1140/epjc/s10052-019-7452-6}
  {\bibfield  {journal} {\bibinfo  {journal} {Eur. Phys. J. C}\ }\textbf
  {\bibinfo {volume} {79}},\ \bibinfo {pages} {928} (\bibinfo {year} {2019})},\
  \Eprint {https://arxiv.org/abs/1911.00302} {arXiv:1911.00302 [gr-qc]}
  \BibitemShut {NoStop}%
\bibitem [{\citenamefont {Jiang}\ and\ \citenamefont
  {Zhang}(2020)}]{Jiang:2020btc}%
  \BibitemOpen
  \bibfield  {author} {\bibinfo {author} {\bibfnamefont {J.}~\bibnamefont
  {Jiang}}\ and\ \bibinfo {author} {\bibfnamefont {M.}~\bibnamefont {Zhang}},\
  }\bibfield  {title} {\bibinfo {title} {{Weak cosmic censorship conjecture in
  Einstein\textendash{}Maxwell gravity with scalar hair}},\ }\href
  {https://doi.org/10.1140/epjc/s10052-020-7751-y} {\bibfield  {journal}
  {\bibinfo  {journal} {Eur. Phys. J. C}\ }\textbf {\bibinfo {volume} {80}},\
  \bibinfo {pages} {196} (\bibinfo {year} {2020})}\BibitemShut {NoStop}%
\bibitem [{\citenamefont {Chowdhury}\ and\ \citenamefont
  {Banerjee}(2018)}]{Chowdhury:2018pre}%
  \BibitemOpen
  \bibfield  {author} {\bibinfo {author} {\bibfnamefont {A.}~\bibnamefont
  {Chowdhury}}\ and\ \bibinfo {author} {\bibfnamefont {N.}~\bibnamefont
  {Banerjee}},\ }\bibfield  {title} {\bibinfo {title} {{Quasinormal modes of a
  charged spherical black hole with scalar hair for scalar and Dirac
  perturbations}},\ }\href {https://doi.org/10.1140/epjc/s10052-018-6065-9}
  {\bibfield  {journal} {\bibinfo  {journal} {Eur. Phys. J. C}\ }\textbf
  {\bibinfo {volume} {78}},\ \bibinfo {pages} {594} (\bibinfo {year} {2018})},\
  \Eprint {https://arxiv.org/abs/1807.09559} {arXiv:1807.09559 [gr-qc]}
  \BibitemShut {NoStop}%
\bibitem [{\citenamefont {Zou}\ and\ \citenamefont
  {Myung}(2020{\natexlab{b}})}]{Zou:2019ays}%
  \BibitemOpen
  \bibfield  {author} {\bibinfo {author} {\bibfnamefont {D.-C.}\ \bibnamefont
  {Zou}}\ and\ \bibinfo {author} {\bibfnamefont {Y.~S.}\ \bibnamefont
  {Myung}},\ }\bibfield  {title} {\bibinfo {title} {{Scalar hairy black holes
  in Einstein-Maxwell-conformally coupled scalar theory}},\ }\href
  {https://doi.org/10.1016/j.physletb.2020.135332} {\bibfield  {journal}
  {\bibinfo  {journal} {Phys. Lett. B}\ }\textbf {\bibinfo {volume} {803}},\
  \bibinfo {pages} {135332} (\bibinfo {year} {2020}{\natexlab{b}})},\ \Eprint
  {https://arxiv.org/abs/1911.08062} {arXiv:1911.08062 [gr-qc]} \BibitemShut
  {NoStop}%
\bibitem [{\citenamefont {Cunha}\ \emph {et~al.}(2016)\citenamefont {Cunha},
  \citenamefont {Herdeiro}, \citenamefont {Radu},\ and\ \citenamefont
  {Runarsson}}]{Cunha:2016bpi}%
  \BibitemOpen
  \bibfield  {author} {\bibinfo {author} {\bibfnamefont {P.~V.~P.}\
  \bibnamefont {Cunha}}, \bibinfo {author} {\bibfnamefont {C.~A.~R.}\
  \bibnamefont {Herdeiro}}, \bibinfo {author} {\bibfnamefont {E.}~\bibnamefont
  {Radu}},\ and\ \bibinfo {author} {\bibfnamefont {H.~F.}\ \bibnamefont
  {Runarsson}},\ }\bibfield  {title} {\bibinfo {title} {{Shadows of Kerr black
  holes with and without scalar hair}},\ }\href
  {https://doi.org/10.1142/S0218271816410212} {\bibfield  {journal} {\bibinfo
  {journal} {Int. J. Mod. Phys. D}\ }\textbf {\bibinfo {volume} {25}},\
  \bibinfo {pages} {1641021} (\bibinfo {year} {2016})},\ \Eprint
  {https://arxiv.org/abs/1605.08293} {arXiv:1605.08293 [gr-qc]} \BibitemShut
  {NoStop}%
\bibitem [{\citenamefont {Kosyakov}(2007)}]{kosyakov2007introduction}%
  \BibitemOpen
  \bibfield  {author} {\bibinfo {author} {\bibfnamefont {B.}~\bibnamefont
  {Kosyakov}},\ }\href@noop {} {\emph {\bibinfo {title} {Introduction to the
  classical theory of particles and fields}}}\ (\bibinfo  {publisher} {Springer
  Science \& Business Media},\ \bibinfo {year} {2007})\BibitemShut {NoStop}%
\bibitem [{\citenamefont {Mann}\ \emph {et~al.}(2021)\citenamefont {Mann},
  \citenamefont {Pando~Zayas},\ and\ \citenamefont {Park}}]{Mann:2020wad}%
  \BibitemOpen
  \bibfield  {author} {\bibinfo {author} {\bibfnamefont {R.~B.}\ \bibnamefont
  {Mann}}, \bibinfo {author} {\bibfnamefont {L.~A.}\ \bibnamefont
  {Pando~Zayas}},\ and\ \bibinfo {author} {\bibfnamefont {M.}~\bibnamefont
  {Park}},\ }\bibfield  {title} {\bibinfo {title} {{Complement to
  thermodynamics of dyonic Taub-NUT-AdS spacetime}},\ }\href
  {https://doi.org/10.1007/JHEP03(2021)039} {\bibfield  {journal} {\bibinfo
  {journal} {JHEP}\ }\textbf {\bibinfo {volume} {03}},\ \bibinfo {pages}
  {039}},\ \Eprint {https://arxiv.org/abs/2012.13506} {arXiv:2012.13506
  [hep-th]} \BibitemShut {NoStop}%
\bibitem [{\citenamefont {Abbasvandi}\ \emph {et~al.}(2021)\citenamefont
  {Abbasvandi}, \citenamefont {Tavakoli},\ and\ \citenamefont
  {Mann}}]{Abbasvandi:2021nyv}%
  \BibitemOpen
  \bibfield  {author} {\bibinfo {author} {\bibfnamefont {N.}~\bibnamefont
  {Abbasvandi}}, \bibinfo {author} {\bibfnamefont {M.}~\bibnamefont
  {Tavakoli}},\ and\ \bibinfo {author} {\bibfnamefont {R.~B.}\ \bibnamefont
  {Mann}},\ }\bibfield  {title} {\bibinfo {title} {{Thermodynamics of Dyonic
  NUT Charged Black Holes with Entropy as Noether Charge}},\ }\href@noop {} {\
  (\bibinfo {year} {2021})},\ \Eprint {https://arxiv.org/abs/2107.00182}
  {arXiv:2107.00182 [hep-th]} \BibitemShut {NoStop}%
\bibitem [{\citenamefont {Hennigar}\ \emph {et~al.}(2019)\citenamefont
  {Hennigar}, \citenamefont {Kubiz\v{n}\'ak},\ and\ \citenamefont
  {Mann}}]{Hennigar:2019ive}%
  \BibitemOpen
  \bibfield  {author} {\bibinfo {author} {\bibfnamefont {R.~A.}\ \bibnamefont
  {Hennigar}}, \bibinfo {author} {\bibfnamefont {D.}~\bibnamefont
  {Kubiz\v{n}\'ak}},\ and\ \bibinfo {author} {\bibfnamefont {R.~B.}\
  \bibnamefont {Mann}},\ }\bibfield  {title} {\bibinfo {title} {{Thermodynamics
  of Lorentzian Taub-NUT spacetimes}},\ }\href
  {https://doi.org/10.1103/PhysRevD.100.064055} {\bibfield  {journal} {\bibinfo
   {journal} {Phys. Rev. D}\ }\textbf {\bibinfo {volume} {100}},\ \bibinfo
  {pages} {064055} (\bibinfo {year} {2019})},\ \Eprint
  {https://arxiv.org/abs/1903.08668} {arXiv:1903.08668 [hep-th]} \BibitemShut
  {NoStop}%
\bibitem [{\citenamefont {Bordo}\ \emph {et~al.}(2019)\citenamefont {Bordo},
  \citenamefont {Gray},\ and\ \citenamefont {Kubiz\v{n}\'ak}}]{Bordo:2019slw}%
  \BibitemOpen
  \bibfield  {author} {\bibinfo {author} {\bibfnamefont {A.~B.}\ \bibnamefont
  {Bordo}}, \bibinfo {author} {\bibfnamefont {F.}~\bibnamefont {Gray}},\ and\
  \bibinfo {author} {\bibfnamefont {D.}~\bibnamefont {Kubiz\v{n}\'ak}},\
  }\bibfield  {title} {\bibinfo {title} {{Thermodynamics and Phase Transitions
  of NUTty Dyons}},\ }\href {https://doi.org/10.1007/JHEP07(2019)119}
  {\bibfield  {journal} {\bibinfo  {journal} {JHEP}\ }\textbf {\bibinfo
  {volume} {07}},\ \bibinfo {pages} {119}},\ \Eprint
  {https://arxiv.org/abs/1904.00030} {arXiv:1904.00030 [hep-th]} \BibitemShut
  {NoStop}%
\bibitem [{\citenamefont {Ballon~Bordo}\ \emph {et~al.}(2020)\citenamefont
  {Ballon~Bordo}, \citenamefont {Gray},\ and\ \citenamefont
  {Kubiz\v{n}\'ak}}]{BallonBordo:2020mcs}%
  \BibitemOpen
  \bibfield  {author} {\bibinfo {author} {\bibfnamefont {A.}~\bibnamefont
  {Ballon~Bordo}}, \bibinfo {author} {\bibfnamefont {F.}~\bibnamefont {Gray}},\
  and\ \bibinfo {author} {\bibfnamefont {D.}~\bibnamefont {Kubiz\v{n}\'ak}},\
  }\bibfield  {title} {\bibinfo {title} {{Thermodynamics of Rotating NUTty
  Dyons}},\ }\href {https://doi.org/10.1007/JHEP05(2020)084} {\bibfield
  {journal} {\bibinfo  {journal} {JHEP}\ }\textbf {\bibinfo {volume} {05}},\
  \bibinfo {pages} {084}},\ \Eprint {https://arxiv.org/abs/2003.02268}
  {arXiv:2003.02268 [hep-th]} \BibitemShut {NoStop}%
\bibitem [{\citenamefont {Ballon~Bordo}\ \emph {et~al.}(2019)\citenamefont
  {Ballon~Bordo}, \citenamefont {Gray}, \citenamefont {Hennigar},\ and\
  \citenamefont {Kubiz\v{n}\'ak}}]{BallonBordo:2019vrn}%
  \BibitemOpen
  \bibfield  {author} {\bibinfo {author} {\bibfnamefont {A.}~\bibnamefont
  {Ballon~Bordo}}, \bibinfo {author} {\bibfnamefont {F.}~\bibnamefont {Gray}},
  \bibinfo {author} {\bibfnamefont {R.~A.}\ \bibnamefont {Hennigar}},\ and\
  \bibinfo {author} {\bibfnamefont {D.}~\bibnamefont {Kubiz\v{n}\'ak}},\
  }\bibfield  {title} {\bibinfo {title} {{The First Law for Rotating NUTs}},\
  }\href {https://doi.org/10.1016/j.physletb.2019.134972} {\bibfield  {journal}
  {\bibinfo  {journal} {Phys. Lett. B}\ }\textbf {\bibinfo {volume} {798}},\
  \bibinfo {pages} {134972} (\bibinfo {year} {2019})},\ \Eprint
  {https://arxiv.org/abs/1905.06350} {arXiv:1905.06350 [hep-th]} \BibitemShut
  {NoStop}%
\bibitem [{\citenamefont {Awad}\ and\ \citenamefont
  {Eissa}(2020)}]{Awad:2020dhy}%
  \BibitemOpen
  \bibfield  {author} {\bibinfo {author} {\bibfnamefont {A.}~\bibnamefont
  {Awad}}\ and\ \bibinfo {author} {\bibfnamefont {S.}~\bibnamefont {Eissa}},\
  }\bibfield  {title} {\bibinfo {title} {{Topological dyonic Taub-Bolt/NUT-AdS
  solutions: Thermodynamics and first law}},\ }\href
  {https://doi.org/10.1103/PhysRevD.101.124011} {\bibfield  {journal} {\bibinfo
   {journal} {Phys. Rev. D}\ }\textbf {\bibinfo {volume} {101}},\ \bibinfo
  {pages} {124011} (\bibinfo {year} {2020})},\ \Eprint
  {https://arxiv.org/abs/2007.10489} {arXiv:2007.10489 [gr-qc]} \BibitemShut
  {NoStop}%
\bibitem [{\citenamefont {Ashtekar}\ \emph {et~al.}(2003)\citenamefont
  {Ashtekar}, \citenamefont {Corichi},\ and\ \citenamefont
  {Sudarsky}}]{Ashtekar:2003jh}%
  \BibitemOpen
  \bibfield  {author} {\bibinfo {author} {\bibfnamefont {A.}~\bibnamefont
  {Ashtekar}}, \bibinfo {author} {\bibfnamefont {A.}~\bibnamefont {Corichi}},\
  and\ \bibinfo {author} {\bibfnamefont {D.}~\bibnamefont {Sudarsky}},\
  }\bibfield  {title} {\bibinfo {title} {{Nonminimally coupled scalar fields
  and isolated horizons}},\ }\href
  {https://doi.org/10.1088/0264-9381/20/15/310} {\bibfield  {journal} {\bibinfo
   {journal} {Class. Quant. Grav.}\ }\textbf {\bibinfo {volume} {20}},\
  \bibinfo {pages} {3413} (\bibinfo {year} {2003})},\ \Eprint
  {https://arxiv.org/abs/gr-qc/0305044} {arXiv:gr-qc/0305044} \BibitemShut
  {NoStop}%
\bibitem [{\citenamefont {Lee}\ \emph {et~al.}(2021)\citenamefont {Lee},
  \citenamefont {Richards}, \citenamefont {Stotyn},\ and\ \citenamefont
  {Park}}]{Lee:2018hrd}%
  \BibitemOpen
  \bibfield  {author} {\bibinfo {author} {\bibfnamefont {Y.}~\bibnamefont
  {Lee}}, \bibinfo {author} {\bibfnamefont {M.}~\bibnamefont {Richards}},
  \bibinfo {author} {\bibfnamefont {S.}~\bibnamefont {Stotyn}},\ and\ \bibinfo
  {author} {\bibfnamefont {M.}~\bibnamefont {Park}},\ }\bibfield  {title}
  {\bibinfo {title} {{Quasilocal Smarr relation for an asymptotically flat
  spacetime}},\ }\href {https://doi.org/10.1140/epjc/s10052-021-09112-w}
  {\bibfield  {journal} {\bibinfo  {journal} {Eur. Phys. J. C}\ }\textbf
  {\bibinfo {volume} {81}},\ \bibinfo {pages} {319} (\bibinfo {year} {2021})},\
  \Eprint {https://arxiv.org/abs/1809.07259} {arXiv:1809.07259 [hep-th]}
  \BibitemShut {NoStop}%
\bibitem [{\citenamefont {Bueno}\ \emph {et~al.}(2018)\citenamefont {Bueno},
  \citenamefont {Cano},\ and\ \citenamefont {Ruip\'erez}}]{Bueno:2018xqc}%
  \BibitemOpen
  \bibfield  {author} {\bibinfo {author} {\bibfnamefont {P.}~\bibnamefont
  {Bueno}}, \bibinfo {author} {\bibfnamefont {P.~A.}\ \bibnamefont {Cano}},\
  and\ \bibinfo {author} {\bibfnamefont {A.}~\bibnamefont {Ruip\'erez}},\
  }\bibfield  {title} {\bibinfo {title} {{Holographic studies of Einsteinian
  cubic gravity}},\ }\href {https://doi.org/10.1007/JHEP03(2018)150} {\bibfield
   {journal} {\bibinfo  {journal} {JHEP}\ }\textbf {\bibinfo {volume} {03}},\
  \bibinfo {pages} {150}},\ \Eprint {https://arxiv.org/abs/1802.00018}
  {arXiv:1802.00018 [hep-th]} \BibitemShut {NoStop}%
\bibitem [{\citenamefont {Sebastiani}\ \emph {et~al.}(2018)\citenamefont
  {Sebastiani}, \citenamefont {Vanzo},\ and\ \citenamefont
  {Zerbini}}]{Sebastiani:2017rxr}%
  \BibitemOpen
  \bibfield  {author} {\bibinfo {author} {\bibfnamefont {L.}~\bibnamefont
  {Sebastiani}}, \bibinfo {author} {\bibfnamefont {L.}~\bibnamefont {Vanzo}},\
  and\ \bibinfo {author} {\bibfnamefont {S.}~\bibnamefont {Zerbini}},\
  }\bibfield  {title} {\bibinfo {title} {{Action growth for black holes in
  modified gravity}},\ }\href {https://doi.org/10.1103/PhysRevD.97.044009}
  {\bibfield  {journal} {\bibinfo  {journal} {Phys. Rev. D}\ }\textbf {\bibinfo
  {volume} {97}},\ \bibinfo {pages} {044009} (\bibinfo {year} {2018})},\
  \Eprint {https://arxiv.org/abs/1710.05686} {arXiv:1710.05686 [hep-th]}
  \BibitemShut {NoStop}%
\bibitem [{\citenamefont {Monteiro}\ \emph {et~al.}(2009)\citenamefont
  {Monteiro}, \citenamefont {Perry},\ and\ \citenamefont
  {Santos}}]{Monteiro:2009tc}%
  \BibitemOpen
  \bibfield  {author} {\bibinfo {author} {\bibfnamefont {R.}~\bibnamefont
  {Monteiro}}, \bibinfo {author} {\bibfnamefont {M.~J.}\ \bibnamefont
  {Perry}},\ and\ \bibinfo {author} {\bibfnamefont {J.~E.}\ \bibnamefont
  {Santos}},\ }\bibfield  {title} {\bibinfo {title} {{Thermodynamic instability
  of rotating black holes}},\ }\href
  {https://doi.org/10.1103/PhysRevD.80.024041} {\bibfield  {journal} {\bibinfo
  {journal} {Phys. Rev. D}\ }\textbf {\bibinfo {volume} {80}},\ \bibinfo
  {pages} {024041} (\bibinfo {year} {2009})},\ \Eprint
  {https://arxiv.org/abs/0903.3256} {arXiv:0903.3256 [gr-qc]} \BibitemShut
  {NoStop}%
\bibitem [{\citenamefont {Durka}(2019)}]{Durka:2019ajz}%
  \BibitemOpen
  \bibfield  {author} {\bibinfo {author} {\bibfnamefont {R.}~\bibnamefont
  {Durka}},\ }\bibfield  {title} {\bibinfo {title} {{The first law of black
  hole thermodynamics for Taub-NUT spacetime}},\ }\href@noop {} {\  (\bibinfo
  {year} {2019})},\ \Eprint {https://arxiv.org/abs/1908.04238}
  {arXiv:1908.04238 [gr-qc]} \BibitemShut {NoStop}%
\bibitem [{\citenamefont {Cl\'ement}\ and\ \citenamefont
  {Gal'tsov}(2020)}]{Clement:2019ghi}%
  \BibitemOpen
  \bibfield  {author} {\bibinfo {author} {\bibfnamefont {G.}~\bibnamefont
  {Cl\'ement}}\ and\ \bibinfo {author} {\bibfnamefont {D.}~\bibnamefont
  {Gal'tsov}},\ }\bibfield  {title} {\bibinfo {title} {{On the Smarr formulas
  for electrovac spacetimes with line singularities}},\ }\href
  {https://doi.org/10.1016/j.physletb.2020.135270} {\bibfield  {journal}
  {\bibinfo  {journal} {Phys. Lett. B}\ }\textbf {\bibinfo {volume} {802}},\
  \bibinfo {pages} {135270} (\bibinfo {year} {2020})},\ \Eprint
  {https://arxiv.org/abs/1908.10617} {arXiv:1908.10617 [gr-qc]} \BibitemShut
  {NoStop}%
\bibitem [{\citenamefont {Carter}(1968)}]{Carter:1968rr}%
  \BibitemOpen
  \bibfield  {author} {\bibinfo {author} {\bibfnamefont {B.}~\bibnamefont
  {Carter}},\ }\bibfield  {title} {\bibinfo {title} {{Global structure of the
  Kerr family of gravitational fields}},\ }\href
  {https://doi.org/10.1103/PhysRev.174.1559} {\bibfield  {journal} {\bibinfo
  {journal} {Phys. Rev.}\ }\textbf {\bibinfo {volume} {174}},\ \bibinfo {pages}
  {1559} (\bibinfo {year} {1968})}\BibitemShut {NoStop}%
\bibitem [{\citenamefont {Cebeci}\ \emph {et~al.}(2016)\citenamefont {Cebeci},
  \citenamefont {\"Ozdemir},\ and\ \citenamefont
  {\c{S}entorun}}]{Cebeci:2015fie}%
  \BibitemOpen
  \bibfield  {author} {\bibinfo {author} {\bibfnamefont {H.}~\bibnamefont
  {Cebeci}}, \bibinfo {author} {\bibfnamefont {N.}~\bibnamefont {\"Ozdemir}},\
  and\ \bibinfo {author} {\bibfnamefont {S.}~\bibnamefont {\c{S}entorun}},\
  }\bibfield  {title} {\bibinfo {title} {{Motion of the charged test particles
  in Kerr-Newman-Taub-NUT spacetime and analytical solutions}},\ }\href
  {https://doi.org/10.1103/PhysRevD.93.104031} {\bibfield  {journal} {\bibinfo
  {journal} {Phys. Rev. D}\ }\textbf {\bibinfo {volume} {93}},\ \bibinfo
  {pages} {104031} (\bibinfo {year} {2016})},\ \Eprint
  {https://arxiv.org/abs/1512.08682} {arXiv:1512.08682 [gr-qc]} \BibitemShut
  {NoStop}%
\bibitem [{\citenamefont {Lim}(2021{\natexlab{a}})}]{Lim:2021ejg}%
  \BibitemOpen
  \bibfield  {author} {\bibinfo {author} {\bibfnamefont {Y.-K.}\ \bibnamefont
  {Lim}},\ }\bibfield  {title} {\bibinfo {title} {{Motion of charged particles
  around a magnetic black hole or topological star with a compact extra
  dimension}},\ }\href {https://doi.org/10.1103/PhysRevD.103.084044} {\bibfield
   {journal} {\bibinfo  {journal} {Phys. Rev. D}\ }\textbf {\bibinfo {volume}
  {103}},\ \bibinfo {pages} {084044} (\bibinfo {year} {2021}{\natexlab{a}})},\
  \Eprint {https://arxiv.org/abs/2102.08531} {arXiv:2102.08531 [gr-qc]}
  \BibitemShut {NoStop}%
\bibitem [{\citenamefont {Lim}(2021{\natexlab{b}})}]{Lim:2020bdj}%
  \BibitemOpen
  \bibfield  {author} {\bibinfo {author} {\bibfnamefont {Y.-K.}\ \bibnamefont
  {Lim}},\ }\bibfield  {title} {\bibinfo {title} {{Null geodesics in the $C$
  metric with a cosmological constant}},\ }\href
  {https://doi.org/10.1103/PhysRevD.103.024007} {\bibfield  {journal} {\bibinfo
   {journal} {Phys. Rev. D}\ }\textbf {\bibinfo {volume} {103}},\ \bibinfo
  {pages} {024007} (\bibinfo {year} {2021}{\natexlab{b}})},\ \Eprint
  {https://arxiv.org/abs/2011.05503} {arXiv:2011.05503 [gr-qc]} \BibitemShut
  {NoStop}%
\bibitem [{\citenamefont {Konoplya}(2019)}]{Konoplya:2019sns}%
  \BibitemOpen
  \bibfield  {author} {\bibinfo {author} {\bibfnamefont {R.~A.}\ \bibnamefont
  {Konoplya}},\ }\bibfield  {title} {\bibinfo {title} {{Shadow of a black hole
  surrounded by dark matter}},\ }\href
  {https://doi.org/10.1016/j.physletb.2019.05.043} {\bibfield  {journal}
  {\bibinfo  {journal} {Phys. Lett. B}\ }\textbf {\bibinfo {volume} {795}},\
  \bibinfo {pages} {1} (\bibinfo {year} {2019})},\ \Eprint
  {https://arxiv.org/abs/1905.00064} {arXiv:1905.00064 [gr-qc]} \BibitemShut
  {NoStop}%
\bibitem [{\citenamefont {Grenzebach}\ \emph {et~al.}(2014)\citenamefont
  {Grenzebach}, \citenamefont {Perlick},\ and\ \citenamefont
  {L\"ammerzahl}}]{Grenzebach:2014fha}%
  \BibitemOpen
  \bibfield  {author} {\bibinfo {author} {\bibfnamefont {A.}~\bibnamefont
  {Grenzebach}}, \bibinfo {author} {\bibfnamefont {V.}~\bibnamefont
  {Perlick}},\ and\ \bibinfo {author} {\bibfnamefont {C.}~\bibnamefont
  {L\"ammerzahl}},\ }\bibfield  {title} {\bibinfo {title} {{Photon Regions and
  Shadows of Kerr-Newman-NUT Black Holes with a Cosmological Constant}},\
  }\href {https://doi.org/10.1103/PhysRevD.89.124004} {\bibfield  {journal}
  {\bibinfo  {journal} {Phys. Rev. D}\ }\textbf {\bibinfo {volume} {89}},\
  \bibinfo {pages} {124004} (\bibinfo {year} {2014})},\ \Eprint
  {https://arxiv.org/abs/1403.5234} {arXiv:1403.5234 [gr-qc]} \BibitemShut
  {NoStop}%
\bibitem [{\citenamefont {Zhang}\ and\ \citenamefont
  {Jiang}(2021)}]{Zhang:2020xub}%
  \BibitemOpen
  \bibfield  {author} {\bibinfo {author} {\bibfnamefont {M.}~\bibnamefont
  {Zhang}}\ and\ \bibinfo {author} {\bibfnamefont {J.}~\bibnamefont {Jiang}},\
  }\bibfield  {title} {\bibinfo {title} {{Shadows of accelerating black
  holes}},\ }\href {https://doi.org/10.1103/PhysRevD.103.025005} {\bibfield
  {journal} {\bibinfo  {journal} {Phys. Rev. D}\ }\textbf {\bibinfo {volume}
  {103}},\ \bibinfo {pages} {025005} (\bibinfo {year} {2021})},\ \Eprint
  {https://arxiv.org/abs/2010.12194} {arXiv:2010.12194 [gr-qc]} \BibitemShut
  {NoStop}%
\bibitem [{\citenamefont {Grenzebach}\ \emph {et~al.}(2015)\citenamefont
  {Grenzebach}, \citenamefont {Perlick},\ and\ \citenamefont
  {L\"ammerzahl}}]{Grenzebach:2015oea}%
  \BibitemOpen
  \bibfield  {author} {\bibinfo {author} {\bibfnamefont {A.}~\bibnamefont
  {Grenzebach}}, \bibinfo {author} {\bibfnamefont {V.}~\bibnamefont
  {Perlick}},\ and\ \bibinfo {author} {\bibfnamefont {C.}~\bibnamefont
  {L\"ammerzahl}},\ }\bibfield  {title} {\bibinfo {title} {{Photon Regions and
  Shadows of Accelerated Black Holes}},\ }\href
  {https://doi.org/10.1142/S0218271815420249} {\bibfield  {journal} {\bibinfo
  {journal} {Int. J. Mod. Phys. D}\ }\textbf {\bibinfo {volume} {24}},\
  \bibinfo {pages} {1542024} (\bibinfo {year} {2015})},\ \Eprint
  {https://arxiv.org/abs/1503.03036} {arXiv:1503.03036 [gr-qc]} \BibitemShut
  {NoStop}%
\bibitem [{\citenamefont {Perlick}\ and\ \citenamefont
  {Tsupko}(2021)}]{Perlick:2021aok}%
  \BibitemOpen
  \bibfield  {author} {\bibinfo {author} {\bibfnamefont {V.}~\bibnamefont
  {Perlick}}\ and\ \bibinfo {author} {\bibfnamefont {O.~Y.}\ \bibnamefont
  {Tsupko}},\ }\bibfield  {title} {\bibinfo {title} {{Calculating black hole
  shadows: review of analytical studies}},\ }\href@noop {} {\  (\bibinfo {year}
  {2021})},\ \Eprint {https://arxiv.org/abs/2105.07101} {arXiv:2105.07101
  [gr-qc]} \BibitemShut {NoStop}%
\bibitem [{\citenamefont {Wei}\ \emph {et~al.}(2019)\citenamefont {Wei},
  \citenamefont {Liu},\ and\ \citenamefont {Mann}}]{Wei:2018xks}%
  \BibitemOpen
  \bibfield  {author} {\bibinfo {author} {\bibfnamefont {S.-W.}\ \bibnamefont
  {Wei}}, \bibinfo {author} {\bibfnamefont {Y.-X.}\ \bibnamefont {Liu}},\ and\
  \bibinfo {author} {\bibfnamefont {R.~B.}\ \bibnamefont {Mann}},\ }\bibfield
  {title} {\bibinfo {title} {{Intrinsic curvature and topology of shadows in
  Kerr spacetime}},\ }\href {https://doi.org/10.1103/PhysRevD.99.041303}
  {\bibfield  {journal} {\bibinfo  {journal} {Phys. Rev. D}\ }\textbf {\bibinfo
  {volume} {99}},\ \bibinfo {pages} {041303} (\bibinfo {year} {2019})},\
  \Eprint {https://arxiv.org/abs/1811.00047} {arXiv:1811.00047 [gr-qc]}
  \BibitemShut {NoStop}%
\bibitem [{\citenamefont {Li}\ \emph {et~al.}(2020)\citenamefont {Li},
  \citenamefont {Guo},\ and\ \citenamefont {Chen}}]{Li:2020drn}%
  \BibitemOpen
  \bibfield  {author} {\bibinfo {author} {\bibfnamefont {P.-C.}\ \bibnamefont
  {Li}}, \bibinfo {author} {\bibfnamefont {M.}~\bibnamefont {Guo}},\ and\
  \bibinfo {author} {\bibfnamefont {B.}~\bibnamefont {Chen}},\ }\bibfield
  {title} {\bibinfo {title} {{Shadow of a Spinning Black Hole in an Expanding
  Universe}},\ }\href {https://doi.org/10.1103/PhysRevD.101.084041} {\bibfield
  {journal} {\bibinfo  {journal} {Phys. Rev. D}\ }\textbf {\bibinfo {volume}
  {101}},\ \bibinfo {pages} {084041} (\bibinfo {year} {2020})},\ \Eprint
  {https://arxiv.org/abs/2001.04231} {arXiv:2001.04231 [gr-qc]} \BibitemShut
  {NoStop}%
\end{thebibliography}
\end{document}